\documentclass[12pt]{article}
\usepackage{latexsym}
\usepackage{amssymb}
\usepackage{amsmath}

\newcommand{\be}{\begin{equation}}
\newcommand{\ee}{\end{equation}}
\def\tr{{\rm tr}}
\newcommand{\Tr}{{\rm Tr}}

\def\cN{{\cal N}}
\def\bea{\begin{eqnarray}}
\def\eea{\end{eqnarray}}
\def\nn{\nonumber}

\begin{document}
\begin{titlepage}

\phantom{x}

\vspace{1cm}

\begin{center}
{\LARGE\bf
Gauge and matter superfield theories on $S^2$
}

\vspace{1.5cm}

{\large\bf I.B.~Samsonov \footnote{On leave from Tomsk Polytechnic
University, 634050 Tomsk, Russia}  and D.~Sorokin}
\\[10pt]
{\it
Dipartimento di Fisica e Astronomia ``Galileo Galilei'', Universit\`a degli Studi di
Padova\\
and\\
INFN, Sezione di Padova, 35131 Padova, Italy}\\[1mm]
{\tt samsonov, sorokin@pd.infn.it}
\end{center}
\vspace{0.5cm}

\begin{abstract}
We develop a superfield formulation of gauge and
matter field theories on a two-dimensional sphere with rigid
$\cN=(2,2)$ as well as extended supersymmetry. The
construction is based on a supercoset  $\frac{SU(2|1)}{U(1)\times U(1)}$ containing
$S^2$ as the bosonic subspace. We derive an explicit form of
supervielbein and covariant derivatives on this coset, and use
them to construct classical superfield actions for gauge and
matter supermultiplets in this superbackground. We then apply
superfield methods for computing one-loop partition functions of
these theories and demonstrate how the localization technique
works directly in the superspace.
\end{abstract}

\end{titlepage}


\numberwithin{equation}{section}

\section{Introduction and summary}
The method of supersymmetric localization has proved to be a very
powerful tool for computing various quantum quantities such as
partition functions, Wilson loops or correlation functions
exactly, at all orders in perturbation theory (see, e.g.,
\cite{Marino,Cremonesi} for reviews). Originally used for
four-dimensional supersymmetric gauge theories in \cite{Pestun},
recently this method has also been applied to study various
non-perturbative aspects of two-dimensional supersymmetric field
theories. In particular, quantum partition functions of
two-dimensional $\cN=(2,2)$ supergauge models on $S^2$ were
computed in \cite{Benini-Cremonesi,Gomes12} and used for studying
Seiberg-like dualities of these models \cite{Gomis14}. Some
mathematical aspects of such dualities were investigated in a
recent paper \cite{Benini14}. In \cite{Kaehler0,Kaehler1,Kaehler2}
it was shown that partition functions of $\cN=(2,2)$ gauge
theories on $S^2$ compute exact K\"ahler potentials
for Calabi-Yau target spaces of $\cN=(2,2)$ non-linear
sigma-models. A systematic construction of
supersymmetric backgrounds as solutions of the $\cN=(2,2)$
supergravity was given recently in \cite{CC}. Some of the above
mentioned results were extended to the two-dimensional manifolds
with boundaries in \cite{bound1,bound2}.

To apply the supersymmetric localization techniques one puts
classical actions for supersymmetric field theories on a compact
manifold with rigid supersymmetry, such as a sphere. A systematic
prescription for constructing such actions was given in \cite{FS}:
 one should couple the gauge and matter
field models to off-shell supergravity theories and then fix the
supergravity background to be e.g.\ a supersymmetric sphere or an
AdS space. In the limit of large Planck mass, the
supergravity fields decouple and one is left with a Lagrangian for
the field theory on a curved background with rigid supersymmetry.
This procedure is equivalent to considering a superfield supergravity
coupled to matter superfields which include all
necessary auxiliary fields. Once the supergravity background is
fixed, one automatically gets superfield theories which respect
all (super)symmetries of the background, see, e.g., \cite{BKbook,KLRST}.
However, off-shell supersymmetry formulations of
supergravity are not always available. Therefore, in some
cases, alternative methods should be used for the construction of actions
for supersymmetric fields on curved backgrounds which do not
require the knowledge of supergravity.

In a recent paper \cite{SS} we applied superfield techniques for
constructing actions for various supersymmetric models on
$S^3$ and computing their partition functions. These superfield models were
formulated on the
supercoset $\frac{SU(2|1)}{U(1)}$ containing $S^3$ as its bosonic
body. The aim of this paper is to introduce, in a similar way, a suitable
curved superspace for supersymmetric gauge and
matter field theories on the two-sphere and develop an approach for studying
their quantum properties directly in the superspace.

The two-dimensional $(2,2)$ superfield supergravity was studied in
a series of papers \cite{Grisaru1,Grisaru2,Grisaru3} and
corresponding matter superfield theories were coupled to $d=2$
supergravity in \cite{Grisaru4}.
Basically, in this paper we take a particular solution of
$\cN=(2,2)$ $d=2$ supergravity corresponding to the Wick rotated
counterpart of $AdS_2$ space, i.e.\ the two-sphere $S^2$, and
consider various matter superfield theories on such a
superbackground. The supergravity solution of our interest is the
supercoset $\frac{SU(2|1)}{U(1)\times U(1)}$ which contains the
two-sphere as its bosonic body. Note that $SU(2|1)$ is the minimal
possible supersymmetry group for the theories on $S^2$ since the
two-component spinors on $S^2$ are complex. In the next section,
we construct the Cartan forms on the supercoset
$\frac{SU(2|1)}{U(1)\times U(1)}$ and use them to define supercovariant
derivatives, supertorsion and supercurvature. These objects
describe the geometry of the background $\cN=(2,2)$ superspace in
which gauge and matter superfields propagate and are used to
construct classical superfield actions on
$\frac{SU(2|1)}{U(1)\times U(1)}$ given in Section 3.

The use of superfields on $\frac{SU(2|1)}{U(1)\times U(1)}$  also allows us to construct Lagrangians for
models with extended supersymmetry. In Section 4 we present
classical actions for an $\cN=(4,4)$ hypermultiplet, and $\cN=(4,4)$ and $\cN=(8,8)$
SYM fields on $S^2$. We also consider theories obtained by the reduction to $S^2$ of the $d=3$ Gaiotto-Witten \cite{GW}
and ABJM \cite{ABJM} models. For all these models
we derive superfield transformations under extended (hidden) supersymmetry which
does not belong to $SU(2|1)$. All
models with extended supersymmetry involve chiral superfields
which can have, in principle, different charges associated to the $U(1)$ R-symmetry generator
of the group $SU(2|1)$. We find constraints on the values of the
R-charges of the chiral superfields imposed by the extended
supersymmetry.

As we will demonstrate, the superfield formulation is
useful not only for constructing classical actions for supersymmetric field
theories on $S^2$, but also for computing their partition
functions. In Section 5 we show how the one-loop partition
functions for gauge and matter superfields computed in
\cite{Benini-Cremonesi,Gomes12} can be derived with the use of the
superspace methods which make  the cancellations between
bosonic and fermionic contributions automatic. Another advantage of the
superfield approach is the trivialization of the procedure of
finding critical points around which the functional integrals
localize. On the Coulomb branch they simply correspond to constant
vacuum values of gauge superfield strengths. In Section 6 we
demonstrate how the standard localization formulas for the partition
functions of $\cN=(2,2)$ gauge theories \cite{Benini-Cremonesi,Gomes12}
appear from functional integrals over gauge superfields on $\frac{SU(2|1)}{U(1)\times U(1)}$.
We apply the localization method for deriving partition functions
of the Gaiotto-Witten and ABJM models reduced to $S^2$.

An important feature of the two-dimensional $(2,2)$
supersymmetric theories is the possibility of having not only conventional chiral and gauge superfields,
but also their twisted counterparts \cite{Gates1984,HPS,RV}.
Quantum partition functions of models on $S^2$ with twisted
supermultiplets were studied in \cite{Kaehler0,Kaehler1}. In the present paper we
restrict ourselves by considering only ordinary $\cN=(2,2)$ multiplets
which have four-dimensional analogs. Superspace study of
partition functions of models with twisted supermultiplets on $S^2$
will be given elsewhere.

We keep the structure of this paper close to the previous one \cite{SS} and use most of the superspace
conventions introduced therein.

\section{ $\frac{SU(2|1)}{U(1)\times U(1)}$ supergeometry}
\subsection{$SU(2|1)$ superalgebra}
The two-dimensional sphere $S^2$ appears as the bosonic body of the
supercoset $\frac{SU(2|1)}{U(1)\times U(1)}$. The superisometry $SU(2|1)$ of this supermanifold
is generated by the Grassmann-even $SU(2)$ generators $J_{\underline
a}=(J_1,J_2,J_2)$ and the $U(1)$ generator $R$, and by the Grassmann-odd supercharges $Q_\alpha$
and $\bar Q_\alpha$ ($\alpha=1,2$). They obey the following
non-zero (anti)commutation relations
\bea
&&
{}[J_{\underline a},J_{\underline b}]=i\varepsilon_{\underline{abc}}J_{\underline{c}}\,,\qquad
[J_{\underline a},Q_\alpha]=-\frac12(\gamma_{\underline a})^\beta_\alpha
Q_\beta\,,
\qquad
[J_{\underline a},\bar Q_\alpha]=-\frac12(\gamma_{\underline a})^\beta_\alpha \bar
Q_\beta\,,
\nn\\&&{}
\{ Q_\alpha ,\bar Q_\beta \}=\gamma^{\underline
a}_{\alpha\beta}J_{\underline{a}}
+\frac12\varepsilon_{\alpha\beta} R\,,\qquad
[R,Q_\alpha]=-Q_\alpha\,,\qquad
[R,\bar Q_\alpha]=\bar Q_\alpha\,.
\label{SU21}
\eea
Here $(\gamma^{\underline a})^\alpha_\beta$ are three-dimensional
gamma-matrices which can be taken to be equal to
the Pauli matrices \footnote{We use the following conventions for the antisymmetric
tensors with vector $\epsilon_{ab}$ and spinor
$\varepsilon_{\alpha\beta}$ indices:
$\epsilon_{12}=\epsilon^{12}=1$,
$\varepsilon_{12}=-\varepsilon^{12}=1$.
The spinor indices are raised and lowered according to the following rules
$\theta_\alpha = \varepsilon_{\alpha\beta}
\theta^\beta$, $\theta^\alpha=\varepsilon^{\alpha\beta}\theta_\beta$ and are contracted as follows
$\theta^2=\theta^\alpha\theta_\alpha$,
$\bar\theta^2=\bar\theta^\alpha\bar\theta_\alpha$.}.

In the $su(2|1)$ superalgebra, $Q_\alpha$ and $\bar Q_\alpha$
are related by the complex conjugation $(Q_\alpha)^*=\varepsilon^{\alpha\beta}\bar Q_\beta$. However, the Wick rotated Lagrangians on $S^2$ are not supposed to be real, so, in general, we will consider $S^2$ (super)fields like $\Phi$ and $\bar\Phi$ as independent
ones. We denote the number of components of the supersymmetry
generators $Q_\alpha$ and $\bar Q_\alpha$ by $\cN=(2,2)$. We
employ this notation to indicate the number of supersymmetries on
$S^2$ by analogy with supersymmetries in 2d spaces of Lorentz signature.

It is convenient to split the $SU(2)$ generators $J_{\underline
a}$ into the $S^2$-boosts $J_a=(J_1,J_2)$ and the $U(1)$-generator $J_3$ and then perform the re-scaling
of the $SU(2|1)$ generators with the $S^2$ radius $r$,
\be
J_a\to P_a=\frac{J_a}{r}\,,\quad J_3\to M=J_3\,,\quad
Q_\alpha \to \sqrt r Q_\alpha\,,\quad
\bar Q_\alpha \to \sqrt r \bar Q_\alpha\,.
\ee
In terms of these generators the (anti)commutation relations of the $su(2|1)$ superalgebra
(\ref{SU21}) take the form
\bea
&&
{}[P_a,P_b]=\frac i{r^2}\epsilon_{ab}M\,,\quad
[M,P_a]=i\epsilon_{ab}P_b\,,\nn\\
&&{}[P_a, Q_\alpha]=-\frac1{2r}(\gamma_a)_\alpha^\beta Q_\beta\,,
\quad
[P_a, \bar Q_\alpha]=-\frac1{2r}(\gamma_a)_\alpha^\beta \bar Q_\beta\,,
\nn\\&&
[M, Q_\alpha]=-\frac1{2}(\gamma_3)_\alpha^\beta Q_\beta\,,
\quad
[M,\bar Q_\alpha]=-\frac1{2}(\gamma_3)_\alpha^\beta \bar Q_\beta\,,
\nn\\
&&
\{ Q_\alpha ,\bar Q_\beta \}=\gamma^a_{\alpha\beta}P_a
+\frac1r\gamma^3_{\alpha\beta} M
+\frac1{2r}\varepsilon_{\alpha\beta} R\,,\quad
[R,Q_\alpha]=-Q_\alpha\,,\quad
[R,\bar Q_\alpha]=\bar Q_\alpha\,.
\label{super-algebra}
\eea
The meaning of this re-scaling is that in the limit $r\to\infty$
the (anti)commutation relations (\ref{super-algebra}) reduce
to the $d=2$ Euclidean flat space superalgebra in which $P_a$ play the
role of the momenta operators while $M$ stands for the angular
momentum.

We will use the $SU(2|1)$ (anti)commutation relations in
the form (\ref{super-algebra}) for constructing the Cartan forms
on the supercoset $\frac{SU(2|1)}{U(1)\times U(1)}$.

\subsection{Supervielbein and $U(1)$-connections}

Let $z^M=(x^m,\theta^\mu,\bar\theta^\mu)$, $m=1,2$, $\mu=1,2$, be local coordinates on
the supercoset $\frac{SU(2|1)}{U(1)\times U(1)}$. In a given coordinate
system the supervielbein on $\frac{SU(2|1)}{U(1)\times U(1)}$ is
described by a set of one-forms
\be
E^A=dz^M E_M{}^A(z)\,,\qquad  E^A=(E^a,E^\alpha,\bar
E^\alpha)\,.
\ee
They are components of the Maurer-Cartan form $J=G^{-1}dG$
\be\label{dG}
J=G^{-1} dG =iE^a P_a+iE^\alpha Q_\alpha +i\bar E^\alpha \bar Q_\alpha
+ i\Omega_{(M)} M+ i\Omega_{(R)} R\,,
\ee
where $G(x,\theta,\bar\theta)$ is a representative of the supercoset
$\frac{SU(2|1)}{U(1)\times U(1)}$, and $\Omega_{(M)}$ and
$\Omega_{(R)}$ are  $U(1)$-connection one-forms corresponding
to the $M$ and $R$ generators of  $SU(2|1)$, respectively.

To find an
explicit form of the supervielbein we consider the
following parametrization of the coset representative
\be\label{bf}
G=b(x)f(\theta,\bar\theta)\,, \quad b(x)=e^{i x^m P_m}
\,,\quad f(\theta,\bar\theta)=e^{i\theta^\alpha Q_\alpha}e^{i\bar\theta^\beta \bar Q_\beta}\,.
\ee
Then
\be\label{dG1}
G^{-1}dG=f^{-1}(d+ie^a(x) P_a+i\omega(x) M)f\,,
\ee
where $e^a(x)=dx^m e^a_m(x)$ and $\omega(x)=dx^m\omega_m(x)$ are bosonic zweibein and the $U(1)$ connection on
$S^2=SU(2)/U(1)$. They obey the torsion-less constraint and
determine the round-sphere curvature
\be
de^a +\omega^{ab}\wedge e^b=0\,,\qquad
d\omega^{ab}=\frac1 {r^2} e^a\wedge e^b\,,
\ee
where $\omega^{ab}=\epsilon^{ab}\omega$. Note that the
indices $a,b,\ldots$ are raised and lowered with the delta-symbol
$\delta_{ab}$ due to the Euclidian signature.

Now, applying the algebra (\ref{super-algebra}) we find the explicit
expressions for the components of the supervielbein and
superconnections in the decomposition (\ref{dG}),
\bea\label{OR}
E^\alpha&=&{\bf d} \theta^\alpha\,,\nn\\
{\bar E}^\alpha&=&{\bf d} \bar\theta^\alpha - \frac1{2r}{\bf d} \theta^\alpha
\bar\theta^2\,,\nn\\
E^a&=&e^a(x)-i{\bf d} \theta^\alpha \bar\theta^\beta
\gamma^a_{\alpha\beta}\,,\nn\\
\Omega_{(M)}&=&\omega(x)-\frac ir {\bf
d} \theta^\alpha \gamma^3_{\alpha\beta}
\bar\theta^\beta\,,\nn\\
\Omega_{(R)}&=&-\frac i{2r}{\bf d} \theta^\alpha\bar\theta_\alpha\,,
\eea
where $\bf d$ is the Killing-spinor covariant differential
\be
{\bf d}  \theta^\alpha= d\theta^\alpha
-\frac i{2r}(\gamma_a)_\beta^\alpha  \theta^\beta e^a
-\frac i2 (\gamma_3)_\beta^\alpha \theta^\beta \omega \,,\qquad
{\bf d}^2=0\,.
\ee
Note that the $\frac{SU(2|1)}{U(1)\times U(1)}$ supergeometry constructed in this
way has a smooth flat limit at $r\rightarrow \infty$.

The inverse supervielbein is given by a set of differential
operators
\be
E_A = E_A{}^M \partial_M\,,\qquad \partial_M=
(\partial_m,\partial_\alpha,\bar\partial_\alpha)
\ee
such that
\be
E_A{}^M E_M{}^B = \delta_A^B\,.
\ee
For instance, in the coordinate system in which the supervielbein $E^A$
is given by (\ref{OR}) we have the following explicit expressions
for the components of $E_A$:
\bea
E_a&=&\partial_a+(\frac i{2r}(\gamma_a)^\alpha_\beta +\frac
i2\omega_a (\gamma_3)^\alpha_\beta)(\theta^\beta\partial_\alpha
+\bar\theta^\beta\bar\partial_\alpha)\,,\nn\\
E_\alpha&=&\partial_\alpha+i\gamma^a_{\alpha\beta}\bar\theta^\beta\partial_a
+[\frac1{2r}(\gamma^a)^\beta_\gamma(\gamma^a)_{\alpha\delta}
+\frac12(\gamma_3)^\beta_\gamma(\gamma^a)_{\alpha\delta}\omega_a]
\theta^\gamma\bar\theta^\delta \partial_\beta
+\frac i4\omega_a \epsilon^{ab}(\gamma^b)^\beta_\alpha
 \bar\theta^2 \bar\partial_\beta\,,\nn\\
\bar E_\alpha&=&\bar\partial_\alpha\,,
\label{inverse-vielbein}
\eea
where $\partial_a = e_a^m (x)\partial_m$  and
$\omega_a=e_a^m\omega_m(x)$ are purely bosonic.

The explicit form of the supervielbein (\ref{OR}) allows us to
find its Berezinian,
\be
E\equiv {\rm Ber}E_M{}^A=\det e_m^a(x) =\sqrt{ h(x)}\,,
\label{ber}
\ee
where $h(x)=\det h_{mn}(x)$ and $h_{mn}(x)$ is a
metric on $S^2$. The Berezinian (\ref{ber}) appears to be
independent of the Grassmann variables in the coordinate system
corresponding to the choice of the coset representative
(\ref{bf}). As a consequence, the supervolume of the coset
$\frac{SU(2|1)}{U(1)\times U(1)}$ vanishes
\be
\int d^2x d^2\theta d^2\bar\theta\,E =0\,.
\label{vanish-volume}
\ee
We stress that this is the coordinate independent property of this supermanifold.

\subsection{Covariant differential, torsion and curvature}
By construction, the differential form (\ref{dG}) obeys the
Maurer-Cartan equation
\be
dJ+J\wedge J=0\,,
\ee
which implies a number of relations for the components of
the supervielbein and superconnections:
\bea
dE^a +\epsilon^{ab}\Omega_{(M)}\wedge E^b
 -i E^\alpha \wedge \bar E^\beta \gamma^a_{\alpha\beta}&=&0\,,
 \label{24a}\nn\\
d\Omega_{(M)}-\frac1{2r^2}E^a\wedge E^b\epsilon_{ab} -\frac ir E^\alpha\wedge \bar E^\beta
 \gamma^3_{\alpha\beta}&=&0\,,\label{24b}\nn\\
d\Omega_{(R)}-\frac i{2r} \varepsilon_{\alpha\beta}E^\alpha\wedge
\bar E^\beta&=&0\,,\label{24c}\nn\\
dE^\alpha - i\Omega_{(R)}\wedge E^\alpha - \frac i{2r} E^a \wedge
E^\beta (\gamma_a)_\beta^\alpha -\frac i{2}\Omega_{(M)}\wedge E^\beta
 (\gamma_3)^\alpha_\beta &=&0\,,\label{24d}\nn\\
d\bar E^\alpha+i\Omega_{(R)}\wedge\bar E^\alpha - \frac i{2r} E^a \wedge
\bar E^\beta(\gamma_a)^\alpha_\beta - \frac i{2} \Omega_{(M)} \wedge
\bar E^\beta(\gamma_3)^\alpha_\beta&=&0\,.
\label{24}
\eea
These equations can be recast in the unified form
\be
{\cal D}E^A = dE^A+\Omega^{AB}\wedge E^B = T^A\,,
\ee
where $T^A$ is the supertorsion with components
\bea
T^a &=& i\gamma^a_{\alpha\beta}E^\alpha \wedge \bar E^\beta\,,\nn\\
T^\alpha &=& \frac i{2r}(\gamma_a)^\alpha_\beta E^a\wedge
E^\beta\,,\nn\\
\bar T^\alpha &=& \frac i{2r}(\gamma_a)^\alpha_\beta E^a\wedge
\bar E^\beta
\label{supertorsion}
\eea
and ${\cal D}=d+\Omega$ is the covariant differential constructed
with the superconnection $\Omega^{AB}$. Non-vanishing components
of the latter are
\bea
\Omega^{ab}&=&\epsilon^{ab}\Omega_{(M)}\,, \nn\\
\Omega^\alpha_{\beta}&=&-i\delta^\alpha_\beta\Omega_{(R)} - \frac i{2}
(\gamma_3)^\alpha_\beta \Omega_{(M)}\,,\nn\\
\bar\Omega^\alpha_{\beta}&=&i\delta^\alpha_\beta \Omega_{(R)} - \frac i{2}
(\gamma_3)^\alpha_\beta \Omega_{(M)}\,.
\eea

Note that the superconnection $\Omega^{AB}$ is Abelian. Hence, the
corresponding supercurvature is simply
\be
{\cal R}^{AB}= d\Omega^{AB}\,,
\ee
or explicitly,
\bea
{\cal R}^{ab}=d\Omega^{ab}&=&\frac1{r^2}E^a\wedge
E^b+\frac i r \epsilon^{ab}\gamma^3_{\alpha\beta}E^\alpha \wedge
\bar E^\beta\,,\nn\\
{\cal R}^\alpha_\beta=d\Omega^\alpha_{\beta}&=&\frac1{2r}\left[\delta^\alpha_\beta \varepsilon_{\gamma\delta}
 +(\gamma^3)^\alpha_\beta (\gamma^3)_{\gamma\delta}\right]
  E^\gamma\wedge \bar E^\delta
  -\frac i{4r^2}(\gamma_3)^\alpha_\beta \epsilon_{ab} E^a\wedge E^b\,,\nn\\
\bar{\cal R}^\alpha_\beta=d\bar\Omega^\alpha_{\beta}&=&\frac1{2r}\left[-\delta^\alpha_\beta \varepsilon_{\gamma\delta}
 +(\gamma^3)^\alpha_\beta
 (\gamma^3)_{\gamma\delta}\right]
  E^\gamma \wedge \bar E^\delta
    -\frac i{4r^2}(\gamma_3)^\alpha_\beta \epsilon_{ab}E^a\wedge
    E^b \,.
\eea
These equations can be rewritten in a compact form
\be
{\cal R}=\frac i{2r^2} M \epsilon_{ab}E^a\wedge E^b
-\left(\frac 1{2r} R \varepsilon_{\alpha\beta} +
\frac 1r M \gamma^3_{\alpha\beta} \right) E^\alpha \wedge \bar
E^\beta\,,
\label{supercurvature}
\ee
where we assume that the angular momentum operator $M$ acts on the tangent
space vectors $v^a$ and spinors $\psi^\alpha$ according to the following rules
\be
M v^a =-i \epsilon^{ab} v^b\,,\qquad
 M \psi^\alpha = -\frac 12(\gamma_3)^\alpha_\beta
 \psi^\beta\,.
\ee
The R-symmetry generator $R$ acts on a complex superfield $\Phi$ carrying the R-charge $q$ as follows
\be
R\Phi = -q \Phi\,, \qquad
R\bar\Phi = q\bar\Phi\,.
\ee

\subsection{Algebra of covariant derivatives}
Let us consider covariant derivatives on the supercoset
$\frac{SU(2|1)}{U(1)\times U(1)}$
\be
{\cal D}_A = E_A + \Omega_A=({\cal D}_a, {\cal D}_\alpha,\bar{\cal
D}_\alpha)
\ee
appearing in the decomposition of the covariant differential ${\cal D}$
\begin{equation}\label{D}
\mathcal D=d+\Omega=E^A\mathcal D_{A}=E^a\mathcal D_a +
E^\alpha\mathcal D_\alpha+\bar E^\alpha\bar{\mathcal  D}_\alpha\,.
\end{equation}
Using the fact that the covariant differential squares to the curvature ${\cal D}^2={\cal
R}$, one gets the following relation for the  covariant derivatives
\be
T^A{\cal D}_A- E^A\wedge E^B{\cal D}_B{\cal D}_A= {\cal R}\,.
\ee
With the use of the explicit expressions for the supertorsion
(\ref{supertorsion}) and curvature (\ref{supercurvature}) we
find the (anti)commutation relations between the covariant derivatives on
$\frac{SU(2|1)}{U(1)\times U(1)}$
\bea&&
[{\cal D}_a,{\cal D}_b]=\frac i{r^2}\epsilon_{ab}M\,,\quad
[{\cal D}_a,{\cal D}_\alpha]=-\frac i{2r}(\gamma_a)_\alpha^\beta
{\cal D}_\beta\,,\quad
[{\cal D}_a,\bar{\cal D}_\alpha]=-\frac i{2r}(\gamma_a)_\alpha^\beta
\bar{\cal D}_\beta\,,\nn\\&&
\{{\cal D}_\alpha, \bar{\cal D}_\beta \}=i\gamma^a_{\alpha\beta}{\cal
D}_a
+\frac 1r \gamma^3_{\alpha\beta}M
+\frac 1{2r}\varepsilon_{\alpha\beta}R\,,\nn\\&&
\{{\cal D}_\alpha,{\cal D}_\beta \}=\{\bar {\cal D}_\alpha,\bar {\cal D}_\beta
\}=0\,.
\label{deriv}
\eea
The generators $M$ and $R$ act on
$\mathcal D_A$ as follows
\bea&&
[M,{\cal D}_a]=-i\epsilon_{ab}{\cal D}_b\,,\quad
[M,{\cal D}_\alpha]=\frac 12(\gamma_3)_\alpha^\beta {\cal D}_\beta\,,\quad
[M,\bar{\cal D}_\alpha]=\frac 12(\gamma_3)_\alpha^\beta \bar{\cal D}_\beta\,,
\nn\\&&
{}[R,{\cal D}_\alpha]={\cal D}_\alpha\,,\qquad
[R,\bar{\cal D}_\alpha]=-\bar{\cal D}_\alpha\,.
\label{deriv1}
\eea

In the coordinate system corresponding to the coset representative (\ref{bf})
the covariant derivatives have the following form
\bea
{\cal D}_a&=&\partial_a+(\frac i{2r}(\gamma_a)^\alpha_\beta +\frac
i2\omega_a (\gamma_3)^\alpha_\beta)(\theta^\beta\partial_\alpha
+\bar\theta^\beta\bar\partial_\alpha)+i\omega_a M\,,\nn\\
{\cal D}_\alpha&=&\partial_\alpha+i\gamma^a_{\alpha\beta}\bar\theta^\beta\partial_a
+[\frac1{2r}(\gamma^a)^\beta_\gamma(\gamma^a)_{\alpha\delta}
+\frac12(\gamma_3)^\beta_\gamma(\gamma^a)_{\alpha\delta}\omega_a]
\theta^\gamma\bar\theta^\delta \partial_\beta
+\frac i4\omega_a \epsilon^{ab}(\gamma^b)^\beta_\alpha
 \bar\theta^2 \bar\partial_\beta
\nn\\&&
-\bar\theta^\beta(\gamma^a_{\alpha\beta}\omega_a-\frac1r
\gamma^3_{\alpha\beta})M +\frac 1{2r} \bar\theta_\alpha R
 \,,\nn\\
\bar {\cal D}_\alpha&=&\bar\partial_\alpha\,.
\label{Dexplicit}
\eea
Here we used the explicit expressions for the superconnection given
in (\ref{OR}) and the inverse
supervielbein (\ref{inverse-vielbein}). One can check that the derivatives
(\ref{Dexplicit}) obey the algebra (\ref{deriv}).

Note that the derivative $\bar {\cal D}_\alpha=\bar\partial_\alpha$ is short in the
coordinates corresponding to the coset representative (\ref{bf}).
Therefore we refer to this coordinate system as the chiral basis. In
principle, one can consider other coordinates, e.g., anti-chiral
in which the derivative ${\cal D}_\alpha$ becomes short or real
coordinates in which the both covariant spinor derivatives have a symmetric form.

\subsection{Killing supervector}
The $SU(2|1)$ transformations of a superfield
$V(z)$ on $\frac{SU(2|1)}{U(1)\times U(1)}$
\be
\delta V = {\mathbb K}V\,,
\label{general-transf}
\ee
are generated by the operator $\mathbb K$ constructed with the use of the
Killing supervector $\xi^A(z)=(\xi^a,\xi^\alpha,\bar\xi^\alpha)$,
\be
{\mathbb K}=\xi^a {\cal D}_a +\xi^\alpha {\cal D}_\alpha +\bar\xi^\alpha
\bar {\cal D}_\alpha
-i\mu(z)M-i\rho(z) R\,.
\label{K}
\ee
Here $\mu(z)$ and $\rho(z)$ are local superfield parameters which are
related to the components of the Killing supervector $\xi^A$ such
that $\mathbb K$ commutes with all the covariant derivatives
\be
[{\mathbb K},{\cal D}_A]=0\,.
\label{eqK}
\ee
Equation (\ref{eqK}) implies a number of
differential equations on the components of the Killing
supervector and the superfunctions $\mu$ and $\rho$
\\
\\
$[{\cal D}_a,{\mathbb K}]=0$ $\Rightarrow$
\begin{subequations}
\bea
&&{\cal D}_{(a}\xi_{b)}=0\label{1a}\,,\\
&&{\cal D}_a \xi_\alpha =\frac
i{2r}(\gamma_a)_{\alpha\beta}\xi^\beta\,,\qquad
{\cal D}_a \bar\xi_\alpha = \frac
i{2r}(\gamma_a)_{\alpha\beta}\bar\xi^\beta\,,\label{1b}\\
&&{\cal D}_a \mu=\frac 1{r^2}\epsilon_{ab}\xi^b\,,
\qquad
{\cal D}_{[a}\xi_{b]}=-\mu\epsilon_{ab}\,,\label{1c}\\
&&{\cal D}_a \rho =0\,.\label{1d}
\eea
\label{1}
\end{subequations}

$[{\cal D}_\alpha,{\mathbb K}]=0$ $\Rightarrow$
\begin{subequations}
\bea
&&{\cal D}_\alpha \xi^a=i\gamma^a_{\alpha\beta}\bar\xi^\beta
\,,\qquad {\cal D}_\alpha\bar\xi^\beta=0\,,\label{2a}\\
&&{\cal D}_{(\alpha}\xi_{\beta)}+\frac i{2r}\xi^a\gamma^a_{\alpha\beta}
+\frac i2\mu \gamma^3_{\alpha\beta}=0\,,\\
&&{\cal D}_\alpha \xi^\alpha+2i\rho=0\,,\\
&&{\cal D}_\alpha \mu-\frac ir
\gamma^3_{\alpha\beta}\bar\xi^\beta=0\,,\\
&&{\cal D}_\alpha \rho-\frac i{2r}\bar\xi_\alpha=0\,.
\eea
\label{2}
\end{subequations}

$[\bar{\cal D}_\alpha,{\mathbb K}]=0$ $\Rightarrow$
\begin{subequations}
\bea
&&\bar{\cal D}_\alpha \xi^a=i\gamma^a_{\alpha\beta}\xi^\beta
\,,\qquad \bar{\cal D}_\alpha\xi^\beta=0\,,\label{3a}\\
&&\bar{\cal D}_{(\alpha}\bar\xi_{\beta)}+\frac i{2r}\xi^a\gamma^a_{\alpha\beta}
+\frac i2\mu \gamma^3_{\alpha\beta}=0\,,\\
&&\bar{\cal D}_\alpha \bar\xi^\alpha-2i\rho=0\,,\\
&&\bar{\cal D}_\alpha \mu-\frac ir
\gamma^3_{\alpha\beta}\xi^\beta=0\,,\\
&&\bar{\cal D}_\alpha \rho+\frac i{2r}\xi_\alpha=0\,.
\eea
\label{3}
\end{subequations}
In particular, (\ref{1a}) and (\ref{1b}) are the Killing vector
and Killing spinor equations, respectively. Eqs.\ (\ref{2a}) and
(\ref{3a}) show that the Killing spinor $\xi^\alpha$ is chiral
while $\bar\xi^\alpha$ is antichiral and they are expressed in
terms of covariant spinor derivatives of the Killing vector. The
other equations allow one to express the superfunctions $\mu(z)$ and
$\rho(z)$ in terms of the Killing vector and spinors. Thus, the
equations (\ref1)--(\ref3) completely define the comonents of the Killing
supervector and the functions $\mu$ and $\rho$ in (\ref K).

The general solution of the equations (\ref1)--(\ref3) has the following form
\bea
\xi^\alpha &=& \bar{\cal D}^2{\cal D}^\alpha \zeta\,,\quad
\bar\xi^\alpha = -{\cal D}^2 \bar {\cal D}^\alpha \zeta\,,\quad
\xi^a =-2i\gamma^a_{\alpha\beta}\bar{\cal D}^\alpha {\cal D}^\beta
 \zeta\,,\nn\\
\mu &=& -\frac{2i}{r}\gamma^3_{\alpha\beta}\bar{\cal D}^\alpha{\cal
D}^\beta \zeta\,,\qquad
\rho=\frac i2 \bar{\cal D}^\alpha {\cal D}_\alpha \zeta\,,
\label{Kil-comp}
\eea
where $\zeta=\zeta(x,\theta,\bar\theta)$ is a covariantly constant
superfield parameter with zero R-charge defined modulo gauge
transformations,
\be
{\cal D}_a \zeta =0\,,\quad
R\zeta =0\,,\quad
\zeta \sim \zeta-i\Lambda +i\bar\Lambda\,.
\label{2.39}
\ee
Here $\Lambda$ is a chiral and covariantly constant superfunction,
$\bar{\cal D}_\alpha\Lambda=0$, ${\cal D}_a \Lambda =0$. Using the
properties (\ref{2.39}) one can check that the superfields
(\ref{Kil-comp}) solve for (\ref1)--(\ref3) and the superfield parameter
$\zeta$ has the number of independent components which are in one-to-one
correspondence with the parameters of the
$SU(2|1)\times U(1)_A$ group where $U(1)_A$ is the group of
external automorphisms of $SU(2|1)$.

As an example, let us consider a chiral superfield $\Phi$, $\bar{\cal D}_\alpha \Phi=0$.
With the use of (\ref{Kil-comp}) its $SU(2|1)$ transformation can
be represented in the following simple form
\bea
\delta \Phi &=&{\mathbb K} \Phi = (\zeta^a {\cal D}_a
+\zeta^\alpha {\cal D}_\alpha -i \mu M -i \rho R )\Phi
\nn\\&=&\bar{\cal D}^2 [({\cal D}^\alpha\zeta)({\cal D}_\alpha
\Phi)]\,.
\label{2.40}
\eea
This formula will be useful in the next sections.

\section{Superfield actions with $(2,2)$ supersymmetry}
The general form of the action for a superfield theory on the supercoset $\frac{SU(2|1)}{U(1)\times U(1)}$ is
\be
S=\int d^6z\,E \,{\cal L}_{\rm f}
+\int d^4z\,{\cal E}\,{\cal L}_{\rm c}
+\int d^4\bar z\,\bar{\cal E}\,\bar{\cal L}_{\rm c}\,,
\label{general-action}
\ee
where ${\cal L}_{\rm f}$ and ${\cal L}_{\rm c}$ are full
and chiral superspace Lagrangians, respectively. The full
superspace  measure $d^6z\,E=d^2x d^2\theta d^2\bar\theta\,E$ and the chiral
one $d^4z\,{\cal E}=d^2x d^2\theta\,{\cal E}$ are
related to each other as follows
\be
d^6z\, E =
-\frac14 d^4z\,{\cal E}\bar{\cal D}^2\,.
\ee
In this section we will construct classical actions
of the form (\ref{general-action}) for gauge and matter
superfields on the supercoset $\frac{SU(2|1)}{U(1)\times U(1)}$.

\subsection{Gauge superfield}
To describe a gauge theory on the supercoset $\frac{SU(2|1)}{U(1)\times U(1)}$ we extend the covariant derivatives ${\cal D}_A$ with gauge
superfield connections $V_A$ which take values in the Lie algebra of
a gauge group $\mathcal G$,
\be
\nabla_A = {\cal D}_A + V_{A}\,,\qquad
V_{A}=(V_a,V_\alpha,\bar V_\alpha)\,.
\ee
Gauge superfield constraints are imposed by requiring that the
gauge-covariant derivatives obey the commutation relations which
correspond to the following deformation of the algebra
(\ref{deriv})
\bea&&
\{\nabla_\alpha,\nabla_\beta \}=\{\bar \nabla_\alpha,\bar \nabla_\beta
\}=0\,,\nn\\&&
\{\nabla_\alpha, \bar\nabla_\beta \}=i\gamma^a_{\alpha\beta}\nabla_a
+\frac 1r \gamma^3_{\alpha\beta}M
+\frac 1{2r}\varepsilon_{\alpha\beta}R+
i\varepsilon_{\alpha\beta}G
+\gamma^3_{\alpha\beta}H\,,\nn\\&&
[\nabla_a,\nabla_b]=\frac i{r^2}\varepsilon_{ab}M+
i{\bf F}_{ab}\,,\nn\\&&
[\nabla_a,\nabla_\alpha]=-\frac i{2r}(\gamma_a)_\alpha^\beta
\nabla_\beta
- (\gamma_a)_\alpha^\beta \bar W_\beta
\,,\quad
[\nabla_a,\bar\nabla_\alpha]=-\frac i{2r}(\gamma_a)_\alpha^\beta
\bar\nabla_\beta
+ (\gamma_a)_\alpha^\beta W_\beta\,,\nn\\&&
[R,\nabla_\alpha]=\nabla_\alpha\,,\quad
[R,\bar\nabla_\alpha]=-\bar\nabla_\alpha\,,\nn\\&&
[M,\nabla_\alpha]=\frac 12(\gamma^3)_\alpha^\beta \nabla_\beta\,,
\quad
[M,\bar\nabla_\alpha]=\frac 12(\gamma^3)_\alpha^\beta
\bar\nabla_\beta\,.
\label{c-deriv}
\eea
Here $G$, $H$, $W_\alpha$, $\bar W_\alpha$ and ${\bf F}_{ab}$ are gauge superfield
strengths subject to Bianchi identities.
In particular, $W_\alpha$ is covariantly chiral while $\bar W_\alpha$
is covariantly anti-chiral,
\be
\bar\nabla_\alpha W_\beta=0\,,\qquad
\nabla_\alpha \bar W_\beta=0\,.
\ee
They satisfy the `standard' Bianchi identity,
\be
\nabla^\alpha W_\alpha=\bar\nabla^\alpha \bar W_\alpha \,.
\ee
The spinorial superfield strengths $W_\alpha$ and $\bar W_\alpha$
are expressed in terms of the scalar superfield strengths $G$ and $H$
\be
\bar W_\alpha = \nabla_\alpha G=-i(\gamma^3)_\alpha^\beta \nabla_\beta H\,,\qquad
W_\alpha=\bar\nabla_\alpha G=i(\gamma^3)_\alpha^\beta \bar\nabla_\beta H\,.
\label{WG}
\ee
$G$ and $H$ are linear superfields
\be
\nabla^2 G= \bar \nabla^2 G=0\,,\qquad
\nabla^2 H= \bar \nabla^2 H=0\,.
\ee

Let us introduce the gauge potential $V$ as
\be
\nabla_\alpha=e^{-V}{\cal D}_\alpha e^V\,,\qquad
\bar \nabla_\alpha = \bar{\cal D}_\alpha\,.
\label{3.9}
\ee
The superfield strengths are expressed in terms of the gauge
superfield $V$ as follows
\bea
G&=& \frac{i}{2}\bar{\cal D}^\alpha(e^{-V}{\cal D}_\alpha
e^V)\,,\qquad
H=-\frac 12\gamma_3^{\alpha\beta}\bar{\cal D}_\alpha
 (e^{-V} {\cal D}_\beta e^V)\,,
\nn\\
W_\alpha&=&-\frac{i}{4}\bar{\cal D}^2(e^{-V}{\cal D}_\alpha e^V)\,,
\quad
\bar W_\alpha=\frac i2\nabla_\alpha \bar{\cal D}^\beta
(e^{-V}{\cal D}_\beta e^V)\,.
\label{GV}
\eea
The gauge transformation for $V$ reads
\be
e^V\longrightarrow e^{i\bar\Lambda} e^V e^{-i\Lambda}\,,
\label{gauge}
\ee
where $\Lambda$ and $\bar\Lambda$ are (anti)chiral,
$\bar{\cal D}_\alpha \Lambda =0$,
${\cal D}_\alpha \bar\Lambda= 0$.
The corresponding gauge transformations for the superfield
strengths (\ref{GV}) are
\be
G\to e^{i\Lambda} G e^{-i\Lambda}\,,\quad
H\to e^{i\Lambda} H e^{-i\Lambda}\,,\quad
W_\alpha \to e^{i\Lambda} W_\alpha e^{-i\Lambda}\,.
\ee

The super Yang-Mills action is given by the integral over the
chiral superspace of the superfield strength $W_\alpha$ squared,
\be
S_{\rm SYM} =\frac2{g^2}\tr\int d^4z \, {\cal E}\,W^\alpha W_\alpha\,,
\label{S-SYM}
\ee
where $g$ is the gauge coupling constant of mass-dimension one,
$[g]=1$. Alternatively, using the identities (\ref{WG})
one can represent (\ref{S-SYM}) as a full superspace
action in the following two equivalent forms
\be
S_{\rm SYM}=-\frac 4{g^2}\tr \int d^6z \,  E\,
G^2=-\frac 4{g^2}\tr \int d^6z \,  E\,
H^2\,.
\label{S-SYM1}
\ee
The variation of the SYM action (\ref{S-SYM}) or (\ref{S-SYM1}) with respect to the gauge potential $V$ has the following form
\be
\delta S_{\rm SYM}=-\frac{4i}{g^2}\tr\int d^6z\,
E \,
\Delta V \nabla^\alpha W_\alpha \,,
\ee
where $\Delta V=e^{-V}\delta e^{V}$ is the gauge-covariant
variation.

The classical SYM action (\ref{S-SYM1}) is a particular case of a
general action for the two superfield strengths $G$ and $H$
\be
\tr\int d^6z \, E \, {\cal H}(G,H)\,,
\ee
where ${\cal H}$ is some function. The action of this form can
appear as part of the low-energy effective action in two-dimensional gauge theories
in the $\cN=(2,2)$ superspace. It would be interesting to
find the explicit form of the function ${\cal H}$ by direct
quantum computations.

Although in supersymmetric two-dimensional gauge theories there is no
Chern-Simons term, one can consider a model which can be obtained by
dimensional reduction of the three-dimensional supersymmetric
Chern-Simons theory to two dimensions. In terms of the gauge
superfields introduced above this action has the
form which is very similar to the $\cN=2$, $d=3$ Chern-Simons action
\cite{Ivanov92}
\be
S_{\rm CS}=i\kappa r\,\tr\int_0^1 dt\int d^6z\,E
\, \bar{\cal D}^\alpha(e^{-tV}{\cal D}_\alpha e^{tV})e^{-tV}\partial_t
e^{tV}\,,
\label{CS}
\ee
where $\kappa$ is a dimensionless coupling constant.
This action has the non-local form because of the integration over the
auxiliary parameter $t$, but its variation is local,
\be
\delta S_{\rm CS}=2\kappa r\,\tr\int d^6z\,E\,G \Delta V\,.
\label{var-BF}
\ee
We stress that in contrast to the three-dimensional gauge theory,
in two dimensions the action (\ref{CS}) is not topological, but
describes the BF-type interaction of component fields (see eq.\ (\ref{CS-comp})).
 For gauge supermultiplet
components this action was considered in \cite{Benini-Cremonesi}.

In (\ref{CS}) the covariant spinor derivatives ${\cal D}_\alpha$ and $\bar{\cal D}_\beta$
 are contracted with the $\varepsilon^{\alpha\beta}$
tensor, however, in two dimensions there is one more invariant
tensor, namely $\gamma^3_{\alpha\beta}$, which can be used for the
contraction of spinor indices. Hence, we can also consider the action
\be
S_{\rm BF}= \kappa r\, \tr \int_0^1 dt\int d^6z\,E(\gamma^3)^{\alpha\beta}
\bar {\cal D}_\alpha(e^{-tV}{\cal D}_\beta e^{tV})
e^{-tV}\partial_t e^{tV}\,,
\label{BF}
\ee
which is supersymmetric and gauge invariant by the same reasoning
as (\ref{CS}). The variation of the action (\ref{BF}) is also
local
\be
\delta S_{\rm BF} = -2\kappa r\,\tr\int d^6z \, E\, H\Delta
V\,.
\ee
In terms of the component fields the action (\ref{BF}) was
considered in \cite{Benini-Cremonesi}.

Finally, we note that the Fayet-Iliopoulos term in the $\cN=(2,2)$
superspace under consideration has the standard form
\be
S_{\rm FI}=4\xi\tr \int d^6z\,E\, V\,,
\label{FI}
\ee
where $\xi$ is a dimensionless coupling constant.

\subsubsection{Component structure}
The $\cN=(2,2)$ vector multiplet consists of two scalars
$\sigma(x)$ and $\eta(x)$, one vector
$A_a(x)=-\frac12\gamma_a^{\alpha\beta}A_{\alpha\beta}$, spinors
$\lambda_\alpha(x)$ and $\bar \lambda_\alpha(x)$ and one auxiliary
field $D(x)$. By dimensional reduction this supermultiplet is related to
the $\cN=1$, $d=4$ vector multiplet. In particular, the scalars
$\sigma$ and $\eta$ originate from the (dimensionally reduced) components of a four-dimensional
vector.

Let us now consider the component structure of the gauge superfield $V$ . The
unphysical components can be eliminated by imposing the
Wess-Zumino gauge
\be
V|=0\,,\quad
{\cal D}_\alpha V|=\bar{\cal D}_\alpha V|=0\,,\quad
{\cal D}^2 V|=\bar{\cal D}^2 V|=0\,,
\ee
while the physical components appear in the following derivatives
of the gauge superfield
\bea
\frac12 [{\cal D}_\alpha,\bar{\cal D}_\beta]V|&=&A_{\alpha\beta}
+\gamma^3_{\alpha\beta}\eta
+i\varepsilon_{\alpha\beta} \sigma\,,\nn\\
-\frac12 \bar{\cal D}^2 {\cal D}_\alpha V|&=&\lambda_\alpha\,,
\quad
-\frac12 {\cal D}^2 \bar{\cal D}_\alpha
V|=\bar\lambda_\alpha\,,\nn\\
\frac1{8}\{ {\cal D}^2,\bar {\cal D}^2  \}V|&=&D\,,
\label{gauge-components}
\eea
where $|$ denotes the component value of a superfield at
$\theta=\bar\theta=0$.

Using the relations (\ref{GV}) we find the component structure
of the superfield strengths:
\bea
G|&=&-\sigma\,,\qquad H| = -\eta\,,\nn\\
W_\alpha |&=&\frac i2 \lambda_\alpha\,,\qquad
\bar W_\alpha |=\frac i2\bar\lambda_\alpha\,,
\nn\\
{\cal D}^\alpha W_\alpha|&=&-iD-\frac{\sigma}{r}\,,\nn\\
{\cal D}_{(\alpha} W_{\beta)}|&=&-\frac i4 \gamma^3_{\alpha\beta}
\epsilon^{ab} F_{ab}
-\frac i2 \hat\nabla_{\alpha\beta} \sigma
+\frac i2 \epsilon^{ab} \gamma^b_{\alpha\beta}\hat\nabla_a \eta
+\frac12\gamma^3_{\alpha\beta}[\eta,\sigma]
-\frac i{2r}\gamma^3_{\alpha\beta}\eta\,,\nn\\
{\cal D}^2 W_\alpha |&=&-\gamma^a_{\alpha\beta}
\hat\nabla_a \bar \lambda^\beta
+[\sigma,\bar\lambda_\alpha]
-i\gamma^3_{\alpha\beta}[\eta,\bar\lambda^\beta]\,,
\label{W-comp}
\eea
where
\bea
\hat\nabla_a \bar\lambda^\beta&=&\hat{\cal D}_a\bar\lambda^\beta
+i[A_a,\bar\lambda^\beta]\,,\nn\\
\hat\nabla_a \sigma &=&\hat{\cal D}_a\sigma + i[A_a,\sigma]\,,
\qquad
(\hat\nabla_{\alpha\beta}=\gamma^a_{\alpha\beta}\hat\nabla_a)\,,\nn\\
F_{ab}&=&\hat{\cal D}_a A_b - \hat{\cal D}_b A_a+i[A_a,A_b]
\eea
and $\hat{\cal D}_a = \partial_a+ i\omega_a (x)M$ is a covariant
derivative on $S^2$.

Using the above relations we get the component structure of the $\cN=(2,2)$ SYM action
(\ref{S-SYM})
\bea
S_{\rm SYM}&=&\frac 2{g^2}  \tr\int d^6z\,{\cal E}\,W^\alpha
W_\alpha
\nn\\&
=&-\frac1{g^2}\tr\int d^2x \sqrt h\left(W^\alpha {\cal D}^2 W_\alpha
-\frac12{\cal D}^\alpha W_\alpha{\cal D}^\beta W_\beta
-{\cal D}_{(\alpha} W_{\beta)}{\cal D}^{(\alpha} W^{\beta)}
\right)\bigg|.
\label{3.26}
\eea
Substituting (\ref{W-comp}) into (\ref{3.26}) we find
\bea
S_{\rm SYM}&=&\frac1{2g^2}\tr\int d^2x\sqrt h\bigg[
V_a V_a + V_3^2 + \left(iD +\frac 1r\sigma\right)^2
\nn\\&&
+i \lambda^\alpha(
\gamma^a_{\alpha\beta}\hat\nabla_a \bar \lambda^\beta
-[\sigma,\bar\lambda_\alpha]+i\gamma^3_{\alpha\beta}
[\eta,\bar\lambda^\beta]
)
\bigg],
\label{S-SYM-comp1}
\eea
where
\be
V_a=\hat\nabla_a \sigma+\epsilon_{ab}\hat\nabla_b \eta\,,\qquad
V_3=\frac12\epsilon^{ab}F_{ab}+i[\eta,\sigma]
+\frac1r \eta\,.
\ee
Since, modulo a total derivative,
\be
V_a V_a +V_3^2 = \left(\frac12 \epsilon^{ab}F_{ab}+\frac1r
\eta\right)^2 + \hat\nabla_a \sigma \hat\nabla_a \sigma
+\hat\nabla_a \eta \hat\nabla_a \eta
-[\eta,\sigma]^2,
\ee
the action (\ref{S-SYM-comp1}) takes the following
equivalent form
\bea
S_{\rm SYM}&=&\frac1{2g^2}\tr\int d^2x\sqrt h\bigg[
\left(\frac12 \epsilon^{ab}F_{ab}+\frac1r
\eta\right)^2 + \hat\nabla_a \sigma \hat\nabla_a \sigma
+ \hat\nabla_a \eta \hat\nabla_a \eta
-[\eta,\sigma]^2 \nn\\&&
+ \left(iD +\frac 1r\sigma\right)^2
+i \lambda^\alpha(
\gamma^a_{\alpha\beta}\hat\nabla_a \bar \lambda^\beta
-[\sigma,\bar\lambda_\alpha]+i\gamma^3_{\alpha\beta}
[\eta,\bar\lambda^\beta]
)
\bigg],
\label{S-SYM-comp2}
\eea

Similarly we find the component structure of the actions
(\ref{CS}), (\ref{BF}) and (\ref{FI}):
\bea
S_{\rm CS}&=&-\frac{i\kappa r}{4}\tr\int d^2x\sqrt h
\left(\eta \epsilon^{ab}F_{ab}
+\bar\lambda^\alpha \lambda_\alpha
-2i\sigma D+\frac{\eta^2}r -\frac{\sigma^2}{r}\right),
\label{CS-comp}\\
S_{\rm BF}&=&\frac{ \kappa r}2 \tr\int d^2x \sqrt h \left(
\eta D -\frac i{r}\eta \sigma
+\frac 12(\gamma^3)^{\alpha\beta}\lambda_\alpha\bar\lambda_\beta
-\frac i2 \epsilon^{ab} F_{ab} \sigma
\right),\label{BF-comp}\\
S_{\rm FI}&=&\xi \tr \int d^2x\sqrt h\, D\,.
\label{FI-comp}
\eea
The actions (\ref{CS-comp}) and (\ref{BF-comp}) were constructed in
\cite{Benini-Cremonesi} for studying partition functions of
supersymmetric gauge theories on $S^2$. In this paper we
gave the superfield forms (\ref{CS}) and (\ref{BF}) of these
actions.

\subsection{Chiral superfield}
The dynamics on $S^2$ of a chiral superfield $\Phi$,
\be
\bar{\cal D}_\alpha \Phi =0 \,,\qquad
{\cal D}_\alpha \bar\Phi=0\,,
\ee
with the R-charge $q$,
\be
R\Phi=-q \Phi\,,\qquad
R\bar\Phi =q\bar\Phi
\ee
is described by the conventional Wess-Zumino action
\be
S_{\rm WZ}=4\int d^6z \,E \,
\bar\Phi\Phi
+2\int d^4z \, {\cal E}\, W(\Phi)
+2\int d^4\bar z \, \bar {\cal E}\, \bar W(\bar\Phi)
\,,
\ee
where $W(\Phi)$ is a chiral potential. Note that though the
R-charge of the chiral superfield is arbitrary, the R-charge of
the chiral potential is fixed $R W= -2 W$ to have the opposite value of the
R-charge of the chiral measure.

The chiral multiplet consists of a complex scalar $\phi$, a spinor
$\psi_\alpha$ and an auxiliary field $F$. These fields appear as
the following components of the chiral superfield:
\be
\begin{array}{lll}
\phi(x)=\Phi| && \bar\phi(x)=\bar\Phi|\\
\psi_\alpha(x)={\cal D}_\alpha \Phi| && \bar\psi_\alpha(x)=\bar{\cal
D}_\alpha \bar\Phi|\\
F(x)=-\frac12{\cal D}^2 \Phi|&\qquad&
 \bar F(x)=-\frac12\bar{\cal D}^2\bar \Phi|\,.
\end{array}
\label{chiral-components}
\ee
With such a definition of the component fields the Wess-Zumino action
on $S^2$ has the conventional component form \cite{Benini-Cremonesi,Gomes12}
\bea
S_{\rm WZ}&=&\int d^2x\,\sqrt h\left(
-\phi \hat{\cal D}^a \hat{\cal D}_a \bar \phi - \frac{q^2-2q}{4r^2}\phi\bar\phi
-i\gamma^a_{\alpha\beta}\psi^\alpha \hat{\cal D}_a \bar\psi^\beta
-\frac{ q}{2r}\psi^\alpha \bar\psi_\alpha
+F\bar F
\right)\nn\\&&
+\int d^2x \sqrt h\,
 \left(W'(\phi)F-\frac12W''(\phi)\psi^\alpha\psi_\alpha+c.c.\right)\,.
\eea

The interaction of the chiral superfield
with the gauge superfield $V$ in the adjoint representation is described by the action
\be
S_{\rm ad}=4\,\tr\int d^6z\, E\, e^{-V}\bar\Phi
e^{V}\Phi\,.
\label{Sad}
\ee
It is straightforward to find the component structure of this
action taking into account the definition of the component fields
(\ref{gauge-components}) and (\ref{chiral-components})
\bea
S_{\rm ad}&=&\tr\int d^2x \sqrt h \bigg[
\hat\nabla^a \bar\phi \hat\nabla_a \phi
-\frac{q^2-2q}{4r^2}\bar\phi\phi + D[\phi,\bar\phi]
-\frac{iq}r\bar\phi[\sigma,\phi]
\nn\\&&
+\bar\phi[\eta,[\eta,\phi]]
+\bar\phi[\sigma,[\sigma,\phi]]
+i\bar\psi^\alpha(\gamma^a)_\alpha^\beta \hat\nabla_a \psi_\beta
-\frac{q}{2r}\bar\psi^\alpha \psi_\alpha
\nn\\&&
-\bar\psi^\alpha(\gamma_3)_\alpha^\beta [\eta ,\psi_\beta]
-i\bar\psi^\alpha[\sigma,\psi_\alpha]
-[\bar\phi,\lambda^\alpha] \psi_\alpha
-\bar\psi^\alpha [\bar\lambda_\alpha,\phi]
+\bar F F
\bigg].
\label{chiral-comp-ad}
\eea
The generalization to any other representation of the gauge group
is straightforward.

\section{Models with extended supersymmetry}
In the previous section we considered  supersymmetric field
theories on $S^2$ with minimal $\cN=(2,2)$ supersymmetry. The
supersymmetries (as well as other isometries of the coset $\frac{SU(2|1)}{U(1)\times U(1)}$) are generated by the operator $\mathbb K$ given in
(\ref{K}).

\if{}
 With the use of
superfields on $SU(2|1)/[U(1)\times U(1)]$ the construction of
classical actions of such models is straightforward since all the
supersymmetries are realized explicitly.
\fi

In this section we consider field theories on $S^2$ with an extended number of
supersymmetries using the $\cN=(2,2)$ superfield formulation. The examples to be discussed include the $\cN=(4,4)$ and $\cN=(8,8)$ SYM models, hypermultiplet and two-dimensional analogs of the Gaiotto-Witten \cite{GW} and ABJM
\cite{ABJM} theories.

Note that the classical actions for field models with $\cN=(4,4)$ supersymmetry on $S^2$
can be, in principle, derived from \cite{N4.1,N4.2,N4.3,N4.4,N4.5,N4.6,N4.7} where the $\cN=(4,4)$
superfield supergravity with matter was studied using various
approaches. Here we avoid the discussion of features of two-dimensional
supergravity theories and construct the superfield actions with the use of algebraical
methods.

In the $\cN=(2,2)$ superfield formulation the extra supersymmetries are realized as transformations that mix
different superfields, e.g., the chiral and vector multiplets.
Such transformations are associated with extra Killing spinors, say $\epsilon_\alpha$, which obey the equation
\be
\hat{\cal D}_a \epsilon_\alpha = \frac i{2r} (\gamma_a)_\alpha^\beta
\epsilon_\beta\,.
\label{KilSpinor}
\ee
Here $\hat{\cal D}_a = \partial_a + i\omega_a(x)M$ is the
covariant derivative on $S^2$. The spinor $\epsilon_\alpha$ appears as a component
of a chiral `superfield' parameter
\be
\Upsilon=a+\theta^\alpha \epsilon_\alpha +\theta^2 b\,, \qquad
\bar{\cal D}_\alpha\Upsilon=0\,,
\label{Upsilon}
\ee
subject to the covariant constancy condition
\be
{\cal D}_a \Upsilon =0\,.
\label{DUpsilon=0}
\ee
Indeed, using the explicit form (\ref{Dexplicit}) of the derivative ${\cal
D}_a$ in chiral coordinates one can check that (\ref{DUpsilon=0})
implies (\ref{KilSpinor}) while the bosonic components $a$ and $b$
are constant, $\partial_a a = \partial_a b =0$. These components
should correspond to parameters of a R-symmetry group in a model with extended supersymmetry.

Recall that there is the sign ambiguity in the definition of the
Killing spinors (\ref{KilSpinor}) such that a spinor $\tilde
\epsilon_\alpha$ obeying the equation
\be
\hat{\cal D}_a \tilde\epsilon_\alpha = -\frac i{2r} (\gamma_a)_\alpha^\beta
\tilde\epsilon_\beta
\label{KilSpinor'}
\ee
is also a Killing spinor. As is pointed out in \cite{SS},
the Killing spinors subject to the Killing spinor equations with different signs
play important role in constructing field
theories with extended supersymmetry on $S^3$ since they are
independent in three dimensions. Field theories on $S^3$ which
involve different numbers of ``positive'' and ``negative'' Killing
spinors are, in general, not equivalent though they respect the
same amount of supersymmetry. In two dimensions, however, such
spinors are not independent. Indeed, given the spinor
$\epsilon_\alpha$ one can construct
\be
\epsilon^L = \frac12({\bf 1}+\gamma^3)\epsilon\,,\qquad
\epsilon^R= \frac12({\bf 1}-\gamma^3)\epsilon\,,
\ee
such that
\be
\epsilon = \epsilon^L + \epsilon^R\,,\qquad
\tilde \epsilon = \epsilon^L - \epsilon^R\,.
\ee
Thus, for the construction of the supersymmetric field models on $S^2$ it is
sufficient to consider only the Killing spinors $\epsilon_\alpha$
obeying (\ref{KilSpinor}).

In what follows we will discuss in details $\cN=(4,4)$ SYM theory on $S^2$, while for the other examples we will present only classical actions and corresponding extended supersymmetry transformations under which these actions are invariant.

\subsection{$\cN=(4,4)$ SYM theory}
To construct the action for the $\cN=(4,4)$ SYM theory on
$S^2$ we will follow the same procedure as we used for the $\cN=4$
SYM model on $S^3$ in \cite{SS}.

The $\cN=(4,4)$ gauge supermultiplet in the $\cN=(2,2)$ superspace is
described by a gauge superfield $V(x,\theta,\bar\theta)$ and a
chiral superfield $\Phi(x,\theta,\bar\theta)$ in the adjoint
representation. The latter can have an arbitrary R-charge $q$
\be
R\bar\Phi = q\bar\Phi\,,\qquad
R\Phi =-q \Phi\,.
\ee
A naive generalization of the flat space action for these
superfields to the coset $\frac{SU(2|1)}{U(1)\times U(1)}$ is
\be
S_0=-\frac4{g^2}\tr\int d^6z\, E (G^2 -
\overline{\it\Phi}{\it\Phi})\,,
\label{Sbare}
\ee
where
\be
\overline{\it \Phi}= e^{-V} \bar\Phi e^V\,,\quad {\it \Phi} = \Phi\,,\qquad
\nabla_\alpha \overline{\it \Phi} =0 \,,\quad
\bar\nabla_\alpha{\it \Phi}=0
\label{337}
\ee
are covariantly (anti)chiral superfields. The action (\ref{Sbare}) is
invariant under standard gauge transformations
\be
\Delta V= i\bar\Lambda - i \Lambda\,,\quad
\delta{\it\Phi} = i[\Lambda,{\it\Phi}]\,,\quad
\delta\overline{\it\Phi} =i[\bar\Lambda,\overline{\it\Phi}]
\ee
with the (anti)chiral superfield parameter $\Lambda$ ($\bar\Lambda$).

We should find transformations of hidden $\cN=(2,2)$ supersymmetry
which mix the superfields $\it\Phi$ and $V$. Such
transformations are generated by the Killing spinors (\ref{KilSpinor})
which enter the chiral superfield parameter $\Upsilon$
given in (\ref{Upsilon}) and subject to (\ref{DUpsilon=0}).
Taking into account that such transformations should preserve
covariant chirality of ${\it\Phi}$ and should close on the $SU(2)$
isometry of $S^2$ and an $R$-symmetry we find the following unique form of these transformations
\be
\Delta_\Upsilon V = i(\Upsilon\overline{\it \Phi} -\bar \Upsilon{\it\Phi})\,,
\quad
\delta_\Upsilon{\it \Phi} =\bar \nabla^\alpha G{\cal D}_\alpha \Upsilon
  + \frac q{2r} G\Upsilon\,,\quad
\delta_\Upsilon \overline{\it\Phi} = -\nabla^\alpha G \bar {\cal D}_\alpha \bar \Upsilon
  - \frac q{2r}G\bar \Upsilon\,,
\label{transf-non-Ab}
\ee
where the R-charge of $\Upsilon$ should be the same as of $\Phi$
\be
R\bar\Upsilon = q\bar\Upsilon\,,\qquad
R\Upsilon = -q\Upsilon\,.
\ee
Indeed, using the algebra of the covariant derivatives (\ref{c-deriv})
and the constraint (\ref{DUpsilon=0}) one can check that
\be\label{nd}
\bar\nabla_\alpha \delta_\Upsilon{\it\Phi}=0\,,\qquad
\nabla_\alpha \delta_\Upsilon \overline{\it\Phi}=0\,.
\ee

The commutator of the two transformations (\ref{transf-non-Ab}) with
superfield parameters $\Upsilon_1$ and $\Upsilon_2$ can be written
as
\bea
{}
[\delta_{\Upsilon_2},\delta_{\Upsilon_1}] {\it\Phi} &=& \bar\nabla^2 [({\cal D}^\alpha\zeta)(\nabla_\alpha{\it\Phi})]\,,\qquad
{}[\delta_{\Upsilon_2},\delta_{\Upsilon_1}] \overline{\it\Phi} = -\nabla^2 [
(\bar {\cal D}^\alpha \zeta)(\bar \nabla_\alpha \overline{\it\Phi})]\,,\nn\\
{}[\delta_{\Upsilon_2} ,\delta_{\Upsilon_1}]G&=&
-2i\gamma^a_{\alpha\beta}(\bar{\cal D}^\alpha {\cal D}^\beta \zeta)
 \nabla_a G
+(\bar{\cal D}^2{\cal D}^\alpha \zeta)\nabla_\alpha G
-({\cal D}^2 \bar{\cal D}^\alpha \zeta)\bar\nabla_\alpha G
\,,
\label{commut}
\eea
where
\be
\zeta = \frac14(\bar\Upsilon_1 \Upsilon_2 - \bar\Upsilon_2
\Upsilon_1)\,.
\ee
The equations (\ref{commut}) show that the commutator of the two
transformations (\ref{transf-non-Ab}) for the chiral superfield
has exactly the form (\ref{2.40}) while for the superfield
strength $G$ it has the general form
(\ref{general-transf}), (\ref{K}) with components of the Killing
supervector given in (\ref{Kil-comp}). Therefore, the commutator of the transformations
(\ref{transf-non-Ab}) closes on the symmetries of the coset $\frac{SU(2|1)}{U(1)\times U(1)}$ and, in particular, the hidden $\cN=(2,2)$ supersymmetry contained in (\ref{transf-non-Ab}) closes on the bosonic symmetries of the coset.

It is a simple exercise to check that the action (\ref{Sbare}) is
not invariant under (\ref{transf-non-Ab}) unless $q= 0$,
\be
\delta_\Upsilon S_{\rm 0}=-\frac{2q}{rg^2}\tr\int d^6z\,E\, G(\bar\Upsilon {\it\Phi}-\Upsilon\overline{\it\Phi})\,.
\ee
However, similar to the $\cN=4$ SYM model on
$S^3$ \cite{SS}, this term is canceled against the variation of the
Chern-Simons-like action (\ref{CS})
\bea
S_{\rm CS}&=&-\frac{q}{rg^2}\tr\int_0^1 dt\int d^6z\,E
\, \bar{\cal D}^\alpha(e^{-tV}{\cal D}_\alpha e^{tV})e^{-tV}\partial_t
e^{tV}\,,\\
\delta_\Upsilon S_{\rm CS}&=&-\frac{2q}{rg^2}\tr\int d^6z \, E\, G(\Upsilon \overline{\it \Phi}-\bar\Upsilon {\it
\Phi})\,.
\eea
We thus find  that for a generic $q$ the classical action for the
$\cN=(4,4)$ SYM on $S^2$ is
\be
S_{\rm SYM}^{\cN=(4,4)}=-\frac4{g^2}\tr\int d^6z\, E\left[G^2-
\overline{\it\Phi}{\it\Phi}
+\frac{q}{4r}
\int_0^1 dt\, \bar{\cal D}^\alpha(e^{-tV}{\cal D}_\alpha e^{tV})e^{-tV}\partial_t e^{tV}
\right]\,.
\label{SN=44}
\ee
Being manifestly invariant under $SU(2|1)$, this action is also
invariant under the transformations (\ref{transf-non-Ab}), $\delta_\Upsilon S_{\rm
SYM}^{\cN=(4,4)}=0$. All together these transformations form  the
supergroup $SU(2|2)\times SU(2)_A$, where $SU(2)_A$ is the group of external automorphisms of $SU(2|2)$, which will manifest itself in the component form of the action.

\subsubsection{Component structure of $\cN=(4,4)$ SYM on $S^2$}
\label{Sec411}
The classical action for the $\cN=(4,4)$ SYM theory on $S^2$ in
terms of $\cN=(2,2)$ superfields is given by (\ref{SN=44}). We
stress that this action is gauge invariant and $\cN=(4,4)$
supersymmetric for any value of $q$. It is interesting to consider the
component structure of this action to find possible constraints on the
parameter $q$.

The Lagrangian of the action (\ref{SN=44}) consist of three parts, namely,
the pure $\cN=(2,2)$ SYM Lagrangian given by $G^2$, the Lagrangian for the chiral
superfield in the adjoint representation of the gauge group given by
$\overline{\it\Phi}{\it\Phi}$ and, the Chern-Simons-like part given by the last
term in (\ref{SN=44}). The component structure of these three
terms are given in (\ref{S-SYM-comp1}), (\ref{chiral-comp-ad}) and
(\ref{CS-comp}), respectively. Putting these expressions together
we have
\bea
S_{\rm SYM}^{\cN=(4,4)}&=&\frac1{g^2}\tr\int d^2x\sqrt h({\cal L}_{\rm bos}+{\cal L}_{\rm
ferm})\,,
\label{4.20}\\
{\cal L}_{\rm
bos}&=&\frac18\left(\epsilon^{ab}F_{ab}+\frac{q+2}{r}\eta \right)^2
+\frac12 \nabla^a \sigma \nabla_a \sigma
+\frac12 \nabla^a \eta \nabla_a \eta
+\nabla^a \bar\phi
\nabla_a \phi
\nn\\&&
+\frac{q(2-q)}{4r^2}\phi\bar\phi
+\frac{q(2-q)}{8r^2}\sigma^2 -\frac{q(q+2)}{8r^2}\eta^2
+\bar FF-\frac12(D')^2
\nn\\&&
+\frac{i(2-3q)}{2r}\sigma [\phi,\bar\phi]
-\frac12[\eta,\sigma]^2
+\frac12[\phi,\bar\phi]^2
-[\sigma,\phi][\sigma,\bar\phi]
-[\eta,\phi][\eta,\bar\phi]
\label{4.21}
\,,\\
{\cal L}_{\rm ferm}&=&
\frac i2 \lambda^\alpha(
\gamma^a_{\alpha\beta}\nabla_a \bar \lambda^\beta
-[\sigma,\bar\lambda_\alpha]+i\gamma^3_{\alpha\beta}
[\eta,\bar\lambda^\beta]
)+\frac{q}{4r}\bar\lambda^\alpha \lambda_\alpha
\nn\\&&
+i\bar\psi^\alpha(\gamma^a)_\alpha^\beta \hat\nabla_a \psi_\beta
-\frac{q}{2r}\bar\psi^\alpha \psi_\alpha
-\bar\psi^\alpha(\gamma_3)_\alpha^\beta [\eta ,\psi_\beta]
\nn\\&&
-i\bar\psi^\alpha[\sigma,\psi_\alpha]
-[\bar\phi,\lambda^\alpha] \psi_\alpha
-\bar\psi^\alpha [\bar\lambda_\alpha,\phi]\,,
\label{4.22}
\eea
where we made the following shift of the auxiliary field $D$
\be
D' = D+\frac{i(q-2)}{2r}\sigma -[\phi,\bar\phi]\,.
\ee

The scalars $\phi$, $\bar\phi$ and $\sigma$ and the
auxiliary fields can be unified into $SU(2)_R$ and $SU(2)_A$
triplets, respectively,
\be
\phi_I=(\phi_1,\phi_2,\phi_3)\,,\qquad
{\rm F}_A=({\rm F}_1,{\rm F}_2,{\rm F}_3)\,,
\ee
where
\bea
&&\phi=\frac1{\sqrt2}(\phi_1-i\phi_2)\,,\quad
\bar\phi=\frac1{\sqrt2}(\phi_1+i\phi_2)\,,\quad
\phi_3 = -\sigma\,,\nn\\
&&
F=\frac1{\sqrt2}({\rm F}_1-i{\rm F}_2)\,, \quad
\bar F= \frac1{\sqrt2}({\rm F}_1+i{\rm F}_2)\,,\quad
{\rm F}_3 = iD'\,.
\eea
Note that the scalar $\eta$ is an $SU(2)_R\times SU(2)_A$ singlet.

The spinor fields are unified into $SU(2)_R$ doublets
$\psi_{i\alpha}$ ($i=1,2$)
\be
\bar\psi^{1\alpha} = \frac i{\sqrt2} \bar\lambda^\alpha\,,\quad
\psi_{1\alpha} = \frac i{\sqrt2} \lambda_\alpha\,,\quad
\psi_{2\alpha} = \bar\psi_\alpha\,,\quad
\bar\psi^{2\alpha} = \psi^\alpha\,.
\ee
Then, the Lagrangians (\ref{4.21}) and (\ref{4.22}) can be recast into the $SU(2)_R\times SU(2)_A$
covariant form
\bea
\label{4.27}
{\cal L}_{\rm
bos}&=&\frac18\left(\epsilon^{ab}F_{ab}+\frac{q+2}{r}\eta
\right)^2
+\frac12 \hat\nabla^a \phi^I \hat\nabla_a \phi_I
+\frac12\hat\nabla^a \eta \hat\nabla_a \eta
\nn\\&&
+\frac{q(2-q)}{8r^2}\phi^I \phi_I
-\frac{q(q+2)}{8r^2}\eta^2
+\frac12{\rm F}^A{\rm F}_A
\nn\\&&
+\frac{2-3q}{12r}\varepsilon^{IJK}\phi_I[\phi_J,\phi_K]
-\frac12[\eta,\phi^I][\eta,\phi_I]
-\frac14[\phi^I,\phi^J][\phi_I,\phi_J]
\,,\\
{\cal L}_{\rm ferm}&=&
i\bar\psi^{i\alpha}(\gamma^a)_\alpha^\beta \hat\nabla_a
\psi_{i\beta}
-\frac q{2r}\bar\psi^{i\alpha}\psi_{i\alpha}
-\bar\psi^{i\alpha}(\gamma_3)_\alpha^\beta [\eta,\psi_{i\beta}]
+i\bar\psi^{i\alpha}(\gamma^I)_i^j[\phi_I,\psi_{j\alpha}] \,,
\eea
where $(\gamma^I)_i^j$ are gamma-matrices corresponding to the
$SU(2)_R$ group.
Thus, we see that the action (\ref{SN=44}) being $\cN=(4,4)$ supersymmetric is
invariant under $SU(2)\sim SO(3)$ isometry of $S^2$ and possesses
$SU(2)_R\times SU(2)_A$ R-symmetry. These transformations form the supergroup
$SU(2|2)\times SU(2)_A$ where $SU(2)_A$ acts as the group of
external automorphisms of $SU(2|2)$.

Note that all the scalar fields in (\ref{4.27}) have a non-negative
mass squared only for
\be
q=0\,.
\ee
Therefore, though the action (\ref{SN=44}) is
$\cN=(4,4)$ supersymmetric for any value of $q$,
its zero value $q=0$ is singled out among others by the requirement  of the absence
of tachyons in the theory.
Recall that for the analogous $\cN=4$ SYM model on $S^3$ the constraint
$q=0$ appeared from somewhat different arguments, namely, that the $d=3$ SYM action, containing a Chern--Simons term should be invariant under large gauge transformations \cite{SS}.

It would be of interest to construct an analog of the action
(\ref{SN=44}) in the $AdS_2$ space and to find constraints on the
value of the R-charge $q$ in that model. The $\cN=4$
SYM action in $AdS_3$ space in terms of $\cN=2$ superfields was
considered in a recent paper \cite{KuzN4}.

\subsection{Hypermultiplet}
The hypermultiplet is described by a pair of chiral superfields
$(X_+,X_-)$, $\bar{\cal D}_\alpha X_\pm =0$, which, in
principle, can have different R-charges,
\be
R \bar X_\pm = q_\pm \bar X_\pm\,,\qquad
R X_\pm = -q_\pm X_\pm\,.
\ee
The interaction of the hypermultiplet with the $\cN=(4,4)$ gauge
multiplet $(V,\Phi)$ is described by the action
\bea
S_{\rm hyp}&=&4\,\tr\int d^6z \, E
\left(\bar X_+ e^V X_+ e^{-V} +   \bar X_- e^{V} X_- e^{-V} \right)
\nn\\&&
-2\sqrt2 i\,\tr\int d^4z \,{\cal E} \,X_+ [ \Phi, X_- ]
+2\sqrt2 i\, \tr\int d^4\bar z\,\bar{\cal E}\, \bar X_+ [\bar \Phi, \bar
X_-]\,.
\label{hyper-action}
\eea
Here we consider the hypermultiplets in the adjoint representation of the gauge group although the
generalization to any other representation is straightforward.

The chiral superfield $\Phi$ has an arbitrary R-charge $q$. However, in view of the presence
of the chiral potential in the second line of (\ref{hyper-action})
this charge is related to the R-charges of the hypermultiplet
\be
q+ q_+ + q_- =2\,.
\ee

It is convenient to introduce covariantly (anti)chiral superfields
\be
\bar{\cal X}_+  = e^{-V} \bar X_+ e^V\,,\quad
{\cal X}_+ = X_+\,,\quad
\bar{\cal X}_- = e^{-V}\bar X_- e^{V}\,,\quad
{\cal X}_- = X_-\,.
\ee
For these superfields, the transformations of the hidden $\cN=(2,2)$ supersymmetry (which is
parametrized by the Killing spinors $\epsilon_\alpha$ entering the chiral
superfield parameter $\Upsilon$ as in (\ref{Upsilon})) are
\be
\delta{\cal X}_\pm=\pm \frac1{2\sqrt2}\bar\nabla^2 (\bar\Upsilon \bar{\cal X}_\mp )\,,\qquad
 \delta \bar{\cal X}_\pm =\mp \frac1{2\sqrt2}\nabla^2 (\Upsilon{\cal X}_\mp
 )\,.
\label{susy-hyper-non-Ab}
\ee
Under these transformations the action (\ref{hyper-action}) varies as follows
\be
\delta S_{\rm hyp}=- \frac{q(2-q-2q_+)}{2\sqrt2r^2}\tr\int d^4z\,{\cal
E}\,\Upsilon {\cal X}_+ {\cal X}_-
+\frac{q(2-q-2q_+)}{2\sqrt2r^2}\tr\int d^4\bar z\,\bar{\cal E}\bar\Upsilon
\bar{\cal X}_+ \bar{\cal X}_-\,.
\ee
This variation vanishes if one of the following conditions is
satisfied
\be
q_+=q_-=1-\frac q2\,,\mbox{ \quad or \quad} q=0\,.
\ee
Note that for $q=0$ the R-charges $q_+$ and $q_-$ are not
necessary equal to each other.

For $q\ne 0$ the R-charges $q_+$ and $q_-$ are equal to each other
and the chiral superfields $ X_+$ and $X_-$ form an
$SU(2)$ doublet
\be
X_i = (X_+ , X_-)\,,\qquad
\bar X^i = (\bar X_+ ,\bar X_-)\,.
\ee
In terms of these superfields the action (\ref{hyper-action}) has the following
compact form
\be
S_{\rm hyp} = 4\tr\int d^6z\, E\,\bar {\cal
X}^i {\cal X}_i
-\sqrt2 i \tr\int d^4z\,{\cal E}\,{\cal X}^i[{\it\Phi},{\cal X}_i]
+\sqrt 2i\tr\int d^4\bar z\,\bar{\cal E}\,
\bar{\cal X}^i[\bar{\it\Phi},\bar{\cal X}_i]\,,
\label{Shyp}
\ee
while the hidden supersymmetry transformations
(\ref{susy-hyper-non-Ab}) simplify to
\be
\delta {\cal X}_i=\frac1{2\sqrt2}\bar\nabla^2(\bar\Upsilon \bar{\cal
X}_i)\,,\qquad
\delta\bar {\cal X}^i=\frac1{2\sqrt2}\nabla^2(\Upsilon{\cal X}^i)\,.
\ee
Here the $SU(2)$ indices $i,j$ are raised and lowered with the
antisymmetric tensor
$\varepsilon_{ij}$, $\varepsilon_{12}=\varepsilon^{21}=1$.

\subsection{$\cN=(8,8)$ SYM}
The $\cN=(8,8)$ gauge multiplet consists of the $\cN=(4,4)$ vector
multiplet $(V,\Phi)$ and a hypermultiplet $(X_+, X_-)$ in the
adjoint representation. The $\cN=(8,8)$ SYM action is described
by the sum of the actions (\ref{SN=44}) and (\ref{Shyp})
\be
S_{\rm SYM}^{\cN=(8,8)} = S^{\cN=(4,4)}_{\rm SYM} + S_{\rm hyp}\,.
\label{4.40}
\ee
Recall that $q$ is the R-charge of the chiral superfield $\Phi$
while $q_\pm$ are charges of the hypermultiplet related to $q$ as
$q_+ = q_- = 1-\frac q2$. For arbitrary value of the charge $q$
the action (\ref{4.40}) has only $\cN=(4,4)$ supersymmetry.
However, for $q=\frac 23$ the R-charges of all three chiral
superfields coincide, $q_\pm = \frac23$. In this case the three
chiral superfields form an $SU(3)$ triplet
\be
\Phi_i = (\Phi,X_+,X_-)\,,\quad
\bar\Phi^i = (\bar\Phi,\bar X_+, \bar X_-)\,,\qquad
R\bar \Phi^i = \frac23 \bar\Phi^i\,.
\ee
The action (\ref{4.40}) can be recast into the following form
\bea
S&=&S_{\rm YM}+S_{\rm CS}+S_{\rm pot}\,,\label{N88}\\
S_{\rm YM}&=&-\frac 4{g^2}\tr\int d^6z\,E(G^2- e^{-V}\bar\Phi^i e^V \Phi_i)\,,\\
S_{\rm CS}&=&-\frac{2}{3rg^2} \tr\int_0^1 dt
\int d^6z\,E \,\bar{\cal D}^\alpha
(e^{-tV}{\cal D}_\alpha e^{tV})
e^{-tV}\partial_t e^{tV}\,,\\
S_{\rm pot}&=&-\frac{i\sqrt2}{3g^2}\tr\int  d^4z\, {\cal E}\,
\varepsilon^{ijk}\Phi_i[\Phi_j , \Phi_k]
+\frac{i\sqrt2}{3g^2}\tr\int  d^4\bar z\,
\bar{\cal E}\,
\varepsilon_{ijk}\bar\Phi^i[\bar\Phi^j , \bar\Phi^k]\,.
\eea
One can check that this action is invariant under the following
transformations of a hidden $\cN=(6,6)$ supersymmetry
\bea
\Delta V &=& i\Upsilon_i\bar{\it \Phi}^i - i \bar \Upsilon^i{\it\Phi}_i\,, \\
\delta{\it \Phi}_i &=&\bar \nabla^\alpha G{\cal D}_\alpha \Upsilon_i
+ \frac1{3r} G\Upsilon_i+\frac1{2\sqrt2}\varepsilon_{ijk}\bar\nabla^2(\bar \Upsilon^j \bar
{\it\Phi}^k)\,,\\
\delta \bar{\it\Phi}^i &=& -\nabla^\alpha G \bar {\cal D}_\alpha \bar \Upsilon^i -
\frac1{3r}G\bar \Upsilon^i -\frac1{2\sqrt2} \varepsilon^{ijk}\nabla^2(\Upsilon_j
{\it\Phi}_k)\,,
\eea
where ${\it\Phi}_i$ and $\overline{\it\Phi}^i$ are covariantly
(anti)chiral superfields which are defined similar to eq. (\ref{337}) and
$\Upsilon_i$ is a triplet of chiral superfield parameters, $\bar{\cal D}_\alpha
\Upsilon_i$, subject to the constraint
\be
{\cal D}_a \Upsilon_i = 0\,.
\ee
In components, the superfield parameter $\Upsilon_i$ contains the Killing spinors
$\epsilon_{i\alpha}$ which, together with $\bar\epsilon^i_\alpha$
appearing in $\bar\Upsilon^i$, are
responsible for the extra $\cN=(6,6)$ supersymmetry on $S^2$. This supersymmetry extends the original manifest $\cN=(2,2)$ supersymmetry to $\cN=(8,8)$.

\subsection{Gaiotto-Witten model reduced to $S^2$}

In three dimensions the Gaiotto-Witten \cite{GW} and ABJM \cite{ABJM}
models are superconformal theories with extended
supersymmetry. They play an important role in the $AdS_4/CFT_3$
correspondence. The superfield action for the
Gaiotto-Witten and ABJM models on $S^3$ were constructed in \cite{SS}.

 Being reduced to two dimensions, these theories
are, of course, not superconformal, but still represent
interesting two-dimensional supersymmetric models with extended
supersymmetry. In particular, in a recent paper \cite{Hosomichi}
a relation among the two-dimensional reduction of the
ABJM theory and the $q$-deformed $\cN=(4,4)$ SYM models in flat space was
studied. In this paper we consider analogous models on
the two-sphere $S^2$.

The Gaiotto-Witten theory is described by two gauge superfields
$V$ and $\tilde V$ corresponding to two different gauge groups and
by two chiral superfields (a hypermultiplet), $X_+$ and $X_-$, in
the bi-fundamental representation. In general, the chiral
superfields can have different R-charges
\be
RX_\pm = -q_\pm X_\pm\,,\qquad R\bar X_\pm = q_\pm \bar
X_\pm\,.
\ee
We find that a two-dimensional counterpart of the Gaiotto-Witten action for
these superfields has the following form
\bea
S_{\rm GW}&=&S_{\rm CS}[V]-S_{\rm CS}[\tilde V]+S_X+S_{\rm FI}\,,
\label{GW}\\
S_X&=&4\,\tr\int d^6z\, E(
\bar X_+ e^{V} X_+ e^{-\tilde V}
+ X_- e^{-V}\bar X_- e^{\tilde V}
)\,,\\
S_{\rm FI}&=&\frac i4 \kappa(q_+-q_-)\tr\int d^6z\, E (V +\tilde V)\,,
\label{GW-FI}
\eea
where the terms $S_{\rm CS}[V]$ and $S_{\rm CS}[\tilde V]$ have
the form (\ref{CS}). This action is
invariant under the following transformations
\bea
\Delta V &=& \bar\Sigma {\cal X}_+{\cal X}_- + \Sigma \bar{\cal X}_- \bar{\cal
X_+}\,,\quad
\Delta \tilde V = \bar \Sigma {\cal X}_- {\cal X}_+ + \Sigma \bar{\cal X}_+ \bar{\cal
X}_-\,,\nn\\
\delta {\cal X}_\pm&=&\pm \bar\nabla^2(\bar \Upsilon \bar{\cal
X}_\mp)\,,\quad
\delta\bar{\cal X}_\pm=\pm\nabla^2 (\Upsilon{\cal X}_\mp)\,,
\label{susy-GW}
\eea
where ${\cal X}_\pm$ and $\bar{\cal X}_\pm$ are covariantly
(anti)chiral superfields,
\be
\bar{\cal X}_+=e^{-\tilde V}\bar X_+ e^{V}\,,\quad
{\cal X}_+= X_+\,,\quad
\bar{\cal X}_- = e^{-V}\bar X_- e^{\tilde V}\,,\quad
{\cal X}_- = X_-\,,
\label{covQ}
\ee
and $\Upsilon$ ($\bar \Upsilon$) are (anti)chiral superfield
parameters subject to the constraint
(\ref{DUpsilon=0}). They contain the Killing spinors $\epsilon_\alpha$ and
$\bar\epsilon_\alpha$ as their components. The superfield parameters $\Sigma$ and $\bar\Sigma$ are
not independent, but are related to $\Upsilon$ and $\bar\Upsilon$
as
\be
{\cal D}_\alpha \Sigma =  \frac{8i}{\kappa r}\bar{\cal D}_\alpha \bar
\Upsilon\,,\qquad
\bar{\cal D}_\alpha \bar \Sigma = \frac{8 i}{\kappa r}
 {\cal D}_\alpha \Upsilon\,.
\ee
These equations define $\Sigma$ and $\bar\Sigma$ in terms of
$\Upsilon$ and $\bar\Upsilon$ in the unique way. For instance, for the
chiral superfield parameter $\Upsilon$ given in the form
(\ref{Upsilon}) we find the following component field
decomposition for $\bar\Sigma$ in the chiral coordinate system
\bea
\bar \Sigma &=& \frac{8i}{\kappa r}\Big(
\frac{q_++q_--2}{4r}\bar\theta^2 a
+\bar\theta^\alpha \epsilon_\alpha
+\frac{q_++q_--2}{4r}\bar\theta^2 \theta^\alpha \epsilon_\alpha
\nn\\&&
+2\theta^\alpha \bar\theta_\alpha b
+ \frac{q_++q_--2}{4r}\theta^2 \bar\theta^2 b
- \frac{4 r}{q_++q_-}b
\Big)\,.
\eea
Note that the R-charges of $\Upsilon$ and $\Sigma$ are expressed
in terms of $q_\pm$ as follows
\be
R \Upsilon = (q_++q_--2)\Upsilon \,,\qquad
R \Sigma = -(q_++q_-)\Sigma\,.
\ee

We point out that the FI-term in (\ref{GW}) drops out for $q_+ = q_-$.
Effectively, it compensates the difference of the R-charges of the
chiral superfields such that the action remains $\cN=(4,4)$
supersymmetric.

\subsection{ABJ(M) theory reduced to $S^2$}
ABJM theory is similar to the Gaiotto-Witten model. It is also
described by two gauge superfields $V$ and $\tilde V$, but it has two copies
of chiral superfields in the bi-fundamental representation,
\be
X_{+i}\,,\quad X_-^i\,,\qquad i=1,2\,.
\ee
A priori, we assume that these superfields have arbitrary R-charges
\be
R X_{+i} = -q_+ X_{+i}\,,\quad
R X_-^i = -q_- X_-^i\,.
\label{R-ABJM}
\ee
The transformations of the hidden $\cN=(4,4)$ supersymmetry are
analogous to those for the ABJM model on $S^3$ \cite{SS}
\bea
\Delta V &=&\frac{8i}{\kappa r}( \bar \Upsilon^i{}_j {\cal X}_{+i}{\cal X}_-^j
+ \Upsilon_i{}^j \bar{\cal X}_{-j}\bar{\cal X}_+^i )\,,\nn\\
\Delta \tilde V&=&\frac{8i}{\kappa r}(\bar \Upsilon^j{}_i{\cal X}_-^i {\cal X}_{+j}
 + \Upsilon_i{}^j\bar{\cal X}_+^i \bar{\cal X}_{-j})\,,\\
\delta{\cal X}_{+i}&=& \bar\nabla^2(\bar \Upsilon_i{}^j \bar{\cal X}_{-j})\,,\qquad
\delta{\cal X}_-^j = -\bar\nabla^2(\bar \Upsilon_i{}^j \bar{\cal
X}_+^i)\,,\\
\delta\bar{\cal X}_+^i&=&\nabla^2(\Upsilon^i{}_j{\cal X}_-^j)\,,\qquad
\delta\bar{\cal X}_{-j}=-\nabla^2(\Upsilon^i{}_j{\cal X}_{+i})\,,
\label{transfABJM}
\eea
where ${\cal X}_{\pm i}$ and $\bar{\cal X}_{\pm i }$ are
covariantly (anti)chiral superfields defined similar to
(\ref{covQ}), and $\Upsilon^i{}_j$ is a quartet of chiral
superfield parameters each of which is constrained by
(\ref{DUpsilon=0}). The anti-chiral superfield parameters are now not
independent. They are expressed in terms of $\Upsilon^i{}_j$
\be
\bar\Upsilon^i{}_j = r {\cal D}^2 \Upsilon^i{}_j\,.
\ee
This equation restricts the number of independent parameters in $\bar\Upsilon^i{}_j$ and
$\Upsilon^i{}_j$ such that they involve four Killing spinors
$(\epsilon^i{}_j)_\alpha$ which, together with the manifest $\cN=(2,2)$ supersymmetry,
form the $\cN=(6,6)$ supersymmetry of the ABJ(M) model reduced to $S^2$.

The action invariant under (\ref{transfABJM}) has the
following form
\bea
\label{S-ABJM}
S_{\rm ABJM}&=&S_{\rm CS}[V]-S_{\rm CS}[\tilde V]+S_X +
S_{\rm pot}+S_{\rm FI}\,,\\
S_X&=&4\tr\int d^6z \,E \left(
\bar X_+^i e^{V}X_{+i} e^{-\tilde V} +X_-^i e^{-V}\bar X_{-i} e^{\tilde V}
\right)\,,\\
S_{\rm pot}&=&-\frac{4}{\kappa} \tr\int d^4z\,{\cal E}\left(X_{+i} X_-^i  X_{+j} X_-^j - X_-^i X_{+i}X_-^j X_{+j}\right)
\nn\\&&
-\frac{4}{\kappa} \tr\int d^4\bar z\,\bar{\cal E}
\left(\bar X_{-i}\bar X_+^i \bar X_{-j}\bar X_+^j - \bar X_+^i\bar X_{-i} \bar X_+^j\bar X_{-j}
\right)\,,\\
S_{\rm FI}&=&\frac i4 \kappa(q_+-q_-)\tr\int d^6z\, E ( V +\tilde
V)\,.
\eea
The presence of the term $S_{\rm pot}$ imposes the constraint on
the R-charges $q_\pm$
\be
q_++q_- =1\,.
\ee
Therefore, only one of them is independent.

Similarly to the Gaiotto-Witten model (\ref{GW}), the action
(\ref{S-ABJM}) has the FI-term which effectively compensates the
difference of the R-charges of the chiral superfields such that it
respects the symmetry (\ref{transfABJM}) for an arbitrary value of
$q_+$. Obviously, for $q_+ = q_- = \frac12$ the FI-term drops out.

\section{One-loop partition functions}
One-loop partition functions in the $\cN=(2,2)$ gauge and matter
models on $S^2$ were computed in \cite{Benini-Cremonesi,Gomes12}
using the component field approach. For supersymmetric field
theories the partition functions are given by the ratio of
determinants of operators of quadratic fluctuations of fermionic and
bosonic fields. As a rule, there are many cancellations among
contributions to these determinants due to supersymmetry, so
the final result usually looks quite simple. As in the case of superfield models on $S^3$ considered
in \cite{SS}, the use of
the superfield approach makes these
cancellations automatic.
In this section we re-derive the results of one-loop partition
functions of the chiral and gauge $\cN=(2,2)$ multiplets on $S^2$
using the superfield methods.
\subsection{Chiral superfield interacting with background gauge superfield}
\subsubsection{Single chiral superfield interacting with Abelian gauge superfield}
Let us consider the model of a chiral superfield $\Phi$ minimally interacting with an Abelian gauge
superfield $V$,
\be
\label{S-2chiral}
S=4\int d^6z \,E \,
\overline\Phi e^{V}\Phi =4\int d^6z \,E \,
\overline{\it\Phi}{\it\Phi} \,,
\ee
where
\be
\overline{\it\Phi}= \overline\Phi e^{V}\,,\qquad
{\it\Phi} = \Phi
\ee
are the covariantly (anti)chiral superfields.
In the one-loop approximation the partition function $Z$ is given
by the exponent of the effective action $\Gamma$, $Z= e^\Gamma$.
The latter is proportional to
the trace of the logarithm of the second variational derivative of the
classical action
\be
\Gamma = -\frac12 \Tr\ln S''.
\ee

In the model (\ref{S-2chiral}) it is more convenient to compute
the variation of the effective action, $\delta\Gamma$, which is
expressed in terms of the effective current $\langle J \rangle$ as follows
\be
\label{varGamma}
\delta\Gamma= \int d^6z\, E\,\delta V
\langle J \rangle\,.
\ee
The effective current $\langle J\rangle$, in its turn, is related to
the Green's function of the chiral superfield
$\langle\overline{\it\Phi}(z){\it\Phi}(z')\rangle$ considered at
coincident superspace points, $\langle\overline{\it\Phi}{\it\Phi}\rangle
\equiv\langle\overline{\it\Phi}(z){\it\Phi}(z')\rangle|_{z=z'}$,
\be
\langle J \rangle= \langle \frac{\delta S}{\delta V}\rangle=4\langle
\overline{\it\Phi}{\it \Phi} \rangle\,.
\label{J}
\ee
In what follows we denote this Green's function as
$\langle\overline{\it\Phi}(z){\it\Phi}(z')\rangle\equiv{\rm
G}_{-+}(z,z')$. It obeys the equation
\be
\bar\nabla^2{\rm G}_{-+}(z,z')=\delta_+(z,z')\,,
\ee
where $\delta_+(z,z')$ is a chiral delta-function ($\bar\nabla_\alpha \delta_+(z,z')=0$),
\be
 \delta_+(z,z')=-\frac14\bar\nabla^2 \delta^6(z,z')\,,\qquad
\delta^6(z,z')=\frac1E
\delta^2(x-x')\delta^2(\theta-\theta')\delta^2(\bar\theta-\bar\theta')\,.
\ee
As a result, to obtain the variation of the effective action
(\ref{varGamma}) we should find the Green's function ${\rm G}_{-+}$ at
coincident superspace points.

The procedure of computing Green's functions of chiral
superfields in four-dimensional superspace was developed in
\cite{KM1,KM2}. Following this procedure, we express
${\rm G}_{-+}$ in terms of the covariantly chiral Green's
function ${\rm G}_+$,
\be
{\rm G}_{-+}(z,z')=-\frac14 \nabla^2{\rm G}_+(z,z')\,,
\label{G+-}
\ee
where ${\rm G}_+$ obeys
\be
\square_+{\rm G}_+(z,z')=-\delta_+(z,z')\,,\qquad
\square_+ \equiv \frac14\bar\nabla^2 \nabla^2\,.
\label{G+eq}
\ee
Using the algebra of covariant derivatives (\ref{c-deriv}) we
find the explicit form of the operator $\square_+$ acting on a
chiral superfield
\be
\square_+=-\nabla^a\nabla_a +\frac1{4r^2}+\left(H+\frac1{r} M\right)^2+\left(G-\frac
i{2r}(R+1)\right)^2
+i(\bar\nabla^\alpha\bar W_\alpha)
+2i W^\alpha \nabla_\alpha\,.
\label{box+}
\ee

Let us take a very particular background gauge superfield $V=V_0$ such
that its superfield strengths $G$ and $H$ are constant while $W_\alpha$
and $\bar W_\alpha$ vanish, namely,
\be
G= -\sigma=cosnt\,,\quad
H= -\eta =const\,,\quad
W_{\alpha}=\bar W_{\alpha}=0\,,
\label{const-back}
\ee
where $\sigma$ and $\eta$ are the scalar fields in the $\cN=(2,2)$ gauge
supermultiplet.
Using the equations (\ref{W-comp}) one can show that this
background corresponds to the following values of the component
fields
\be
\eta=-\frac{n}{2r}=const\,, \quad
F_{12} =\frac{n}{2r^2}\,,\quad
\lambda_\alpha=\bar\lambda_\alpha=0\,,\quad
\sigma=\sigma_0=const\,,\quad
D=\frac ir
\sigma_0\,.
\label{comp-background}
\ee
Here $n$ is integer owing to the quantization of the gauge field flux
$\frac1{2\pi}\int F = n\in {\mathbb Z}$, \cite{GNO}, while
$\sigma_0$ is an arbitrary real number. As a result, this background is
parametrized by the pair of the parameters $(n,\sigma_0)$ which
appear as arguments of the partition function  $Z=Z(n,\sigma_0)$.
Note that exactly this background for the $\cN=(2,2)$ gauge
supermultiplet was considered in \cite{Benini-Cremonesi,Gomes12} in the
application of the localization method to supersymmetric models
on $S^2$.

For the background (\ref{const-back}) the form of the operator (\ref{box+})
acting on the chiral superfields with R-charge $q$ simplifies,
\be
\square_+=-\nabla^a \nabla_a +m^2\,,\qquad m^2
 \equiv G^2+ H^2 +\frac{ i}rG(q-1)+\frac{q(2-q)}{4r^2}\,,
\label{square+}
\ee
where $m$ is the effective mass. Here $\nabla_a$ is the superspace
derivative which includes the gauge field connection $A_a$ with
constant field strength $F_{12} =\frac{n}{2r^2}$. In purely
bosonic case the operator $\nabla^a \nabla_a$ is usually
referred to as the covariant Laplacian on $S^2$ with a monopole gauge
field background \cite{Benini-Cremonesi,Gomes12}.

For the gauge superfield background described above
the chiral Green's function ${\rm G}_+$ (\ref{G+eq}) can be written as
\be
{\rm G}_+(z,z')=-\frac14\bar \nabla^2{\rm G_o}(z,z')=-\frac14 \bar\nabla'^2{\rm G_o}(z,z')\,,
\label{G+Gv}
\ee
where $\bar\nabla'^2$ acts on $z'$ and ${\rm G_o}(z,z')$ solves
\be
\square_{\rm o}{\rm G_o}(z,z')=-\delta^6(z,z')\,,
\qquad
\square_{\rm o}= -\nabla^a \nabla_a + m^2\,.
\ee
The operator $\square_{\rm o}$ has the same expression as
$\square_+$ given in eq.\
(\ref{square+}), but it acts on the superfields defined in the
full superspace rather than on the chiral superfields.
To check that (\ref{G+Gv}) obeys (\ref{G+eq}) one should use the identities
\be
[\nabla^2 , \square_{\rm o}]=[\bar\nabla^2,\square_{\rm o}]=0\,,
\label{nabla-square}
\ee
which hold for the gauge superfield background under consideration.

Combining (\ref{G+-}) with (\ref{G+Gv}) we find
\be\label{GG+-}
{\rm G}_{-+}(z,z')=\frac1{16}\nabla^2 \bar\nabla'^2{\rm G_o}(z,z')
=-\frac1{16}\nabla^2 \bar\nabla'^2 \frac1{-\nabla^a \nabla_a
+m^2}\delta^6(z,z')\,.
\ee
Next, using (\ref{nabla-square}) we commute the operators
$\nabla^2$ and $\bar\nabla'^2$ with $(-\nabla^a \nabla_a
+m^2)^{-1}$ and consider the Green's function \eqref{GG+-} at coincident
superspace points
\be
{\rm G}_{-+}(z,z)=-\frac1{-\nabla^a \nabla_a +m^2}\frac1{16}\nabla^2 \bar\nabla'^2
\delta^6(z,z')|_{z=z'}=-\frac1{\Delta_{S^2}
+m^2}\delta^2(x,x')|_{x=x'}\,.
\ee
Note that to get a non-vanishing
result, all the fermionic components of the superspace
delta-function $\delta^6(z,z')$ should be cancelled by
the operators $\nabla^2$ and $\bar\nabla'^2$. The remaining expression is nothing but the trace of the
inverse of the purely bosonic Laplacian $\Delta_{S^2}$
acting on the scalar fields on the $S^2$-sphere
\be
-\tr\frac1{\Delta_{S^2}
+m^2}\propto -\sum_{j}
\frac{d_j}{\lambda_j+m^2}\,,
\label{sum}
\ee
where $\lambda_j$ are the eigenvalues of the Laplace
operator on $S^2$ in the monopole background and $d_j$ are their degeneracies
\cite{Benini-Cremonesi,Gomes12}
\be
\lambda_j=\frac 1{r^2}j\left(j+1\right)-\frac{n^2}{4r^2}\,,\qquad
d_j=2j+1\,,\quad j=\frac{|n|}2,\frac{|n|}2+1,\frac{|n|}2+2,\ldots
\label{bos-spetr}
\ee
The sum (\ref{sum}) is divergent. Regularizing it in a standard
way, $\sum \frac1n = \zeta(1)=\gamma$, we find
\bea
{\rm G}_{-+}(z,z)&=&\frac1{4\pi}\left(
\psi(\frac12+\frac{| n|}2- \frac12\sqrt{1- 4m^2 r^2+n^2 })+\psi(\frac12+\frac{|n|}2+\frac12\sqrt{1-4m^2 r^2+n^2})
\right)\nn\\
&=&\frac1{4\pi}\left(
\psi(\frac q2+\frac{|n|}2+ir\sigma_0) + \psi(1+\frac{|n|}2-\frac q2-ir\sigma_0)
\right),
\label{G-+=cot}
\eea
where $\psi(z)$ is the digamma function which is related to the Euler gamma function
by $\psi(z)=\Gamma'(z)/\Gamma(z)$.
Here we used the explicit expression for the effective mass squared $m^2$
given in (\ref{square+}) which implies the identity
\be
\sqrt{1-4m^2 r^2+n^2}=1-q-2ir\sigma_0\,.
\ee
As a result, the effective current is
\be
\langle J \rangle =\frac1{\pi}\Big(
\psi(\frac q2+\frac{|n|}2+ir \sigma_0 )
+\psi(1+\frac{|n|}2-\frac q2-ir\sigma_0)
\Big).
\ee
Now we substitute this effective current into the variation of the effective
action (\ref{varGamma}) and perform integrations over the Grassmann and
bosonic superspace coordinates,
\bea
\delta\Gamma&=&\int d^6z \,E\, \langle J \rangle
\delta V=\frac14\int d^2 x\, \sqrt h \langle J \rangle
\delta D=\frac {i}{4r}\int d^2 x\,\sqrt h \langle J \rangle
\delta \sigma_0
\nn\\&=&\frac i{4\pi r}\int d^2 x\,\sqrt h \delta \sigma_0
\Big(
\psi(\frac q2+\frac{|n|}2+ir\sigma_0 )
+\psi(1+\frac{|n|}2-\frac q2-ir\sigma_0)
\Big)
\nn\\&=&i r \delta \sigma_0
\Big(
\psi(\frac q2+\frac{|n|}2+ir \sigma_0 )
+\psi(1+\frac{|n|}2-\frac q2-ir\sigma_0)
\Big).
\label{522}
\eea
The integration over the Grassmann variables in the first line of
(\ref{522}) is similar to the computation of the component form of
the FI-term (\ref{FI-comp}) from the superfield action (\ref{FI}).
Here we also used the relation between the values of the auxiliary
field $D$ and the scalar $\sigma$ for the considered background
(\ref{comp-background}). When passing from the second to the third line
in (\ref{522}) we used the fact that the integrand is independent
of $x$ and the remaining integration is just the volume of $S^2$,
$\int d^2 x \sqrt h ={\rm Vol}(S^2)= 4\pi r^2$.

It is a simple exercise to restore the effective action from its
variation (\ref{522})
\be
\Gamma=\ln \frac{\Gamma(\frac q2+\frac{|n|}2+ir\sigma_0)}{
\Gamma(1+\frac{|n|}2-\frac q2-ir\sigma_0)}\,.
\ee
Thus, the partition function of the chiral multiplet on the
background (\ref{comp-background}) is
\be
Z=e^\Gamma =\frac{\Gamma(\frac q2+\frac{|n|}2+ir\sigma_0)}{
\Gamma(1+\frac{|n|}2-\frac q2-ir\sigma_0)}\,.
\label{Zsingle}
\ee
This partition function was originally computed  in
\cite{Benini-Cremonesi,Gomes12} using the component field formulation
of the model (\ref{S-2chiral}). Note that the component field
computations involve the spectra of both the (bosonic) Laplacian and
the (fermionic) Dirac operator on $S^2$, but most of the eigenvalues
of these operators cancel against each other in the ratio of the
one-loop determinants of quadratic fluctuations of the bosonic and
fermionic modes. In superspace, we obtained the same result
(\ref{Zsingle}) without explicit use of the fermionic spectrum, only the knowledge of
the purely bosonic spectrum (\ref{bos-spetr}) was necessary. With
the use of the superfield Green's functions of the chiral superfields all
cancellations among bosons and fermions become automatic.

\subsubsection{Chiral superfield in adjoint representation}
Consider the model of a chiral superfield $\Phi$ interacting with a
background non-Abelian gauge superfield $V$ in the adjoint representation
(\ref{Sad}). We assume that the gauge group is
$U(N)$ and the background gauge superfield takes values in the Cartan
subalgebra,
\be
V={\rm diag}(V_1,V_2,\ldots ,V_N)\,,
\label{Vdiag}
\ee
where each of the diagonal elements $V_I$ in (\ref{Vdiag}) has
constant superfield strengths,
\bea
W_{I\,\alpha } &=&\bar W_{I\, \alpha}  =0\,,\nn\\
G_I&=&\frac i2 \bar{\cal D}^\alpha {\cal D}_\alpha
V_I=-\sigma_I = const\,,\nn\\
H_I &=& -\frac12 (\gamma^3)^{\alpha\beta}\bar{\cal D}_\alpha {\cal D}_\beta V_I
 = -\eta_I = const\,.
\label{G-const}
\eea
In components, such a background is given by
(\ref{comp-background}), but now we will have a set of $N$ independent
pairs $(n_I, \sigma_I)$ as arguments of the partition function,
$Z=Z(n_I,\sigma_I)$.

The (anti)chiral superfield $\it\Phi$ $(\overline{\it\Phi})$ is a
matrix in the $u(N)$ Lie algebra. It can be expanded in the basis
elements $e_{IJ}=(e_{IJ})_{KL}=\delta_{IK}\delta_{JL}$
\be
{\it\Phi}=\sum_{I\ne J}^N e_{IJ}{\it\Phi}_{IJ}
+\sum_{I=1}^N e_{II}{\it\Phi}_I
\,,\qquad
\overline{\it \Phi}=\sum_{I\ne J}^N e_{JI}\overline{\it
\Phi}_{IJ}+\sum_{I=1}^N e_{II}\overline{\it\Phi}_I\,.
\label{Phi-decomp}
\ee
Note that the superfields ${\it\Phi}_{I}$ and
$\overline{\it\Phi}_I$ in (\ref{Phi-decomp}) correspond to the
diagonal elements of the $u(N)$ matrix. These elements do not interact with
the background gauge superfield (\ref{Vdiag}) and we omit them in
what follows. The off-diagonal elements ${\it\Phi}_{IJ}$ enter the action
(\ref{Sad}) as follows
\be
S_{\rm ad}=\sum_{I\ne J}^N \int d^6z\,E\,
\overline{\it\Phi}_{IJ}{\it\Phi}_{IJ}\,,
\label{S-sum}
\ee
where ${\it\Phi}_{IJ}$ are chiral $\bar{\cal D}_\alpha{\it\Phi}_{IJ}=0$
while $\overline{\it\Phi}_{IJ}$ are covariantly antichiral,
\be
e^{V_J-V_I}{\cal D}_\alpha e^{V_I-V_J}
\overline{\it\Phi}_{IJ}=0
\mbox{ for } I<J\, ,\qquad
e^{V_I-V_J}{\cal D}_\alpha e^{V_J-V_I}
\overline{\it\Phi}_{IJ}=0
\mbox{ for } I>J\,.
\label{cov-chir-varphi}
\ee

Each element in the sum (\ref{S-sum}) has the form
(\ref{S-2chiral}). Hence, the partition function in
the model (\ref{Sad}) is given by the product of the expressions
(\ref{Zsingle})
\be
Z_\Phi =\prod_{I\ne J}^N
\frac{\Gamma(\frac q2+\frac{|n_I-n_J|}2+ir(\sigma_I-\sigma_J))}{
\Gamma(1-\frac q2+\frac{|n_I-n_J|}2-ir(\sigma_I-\sigma_J))}\,.
\label{Zad}
\ee
In a similar way one can find the partition function of the chiral
superfield in an arbitrary representation of the gauge group.

Note that for $q=1$ this partition function trivializes,
\be
Z_\Phi|_{q=1}=1\,.
\label{Z=1}
\ee
This property is similar to the one of the partition function of the chiral
superfield on $S^3$ \cite{KWY1}.

\subsection{$\cN=(2,2)$ SYM partition function}
\label{Sec5.2}
Superfield computation of the partition function of $\cN=2$ SYM
on $S^3$ was carried out in \cite{SS}. Here we repeat basic steps of this
procedure for the case of $\cN=(2,2)$ SYM on $S^2$.

At one-loop order the partition function $Z$ is related to the
effective action $\Gamma$ as follows
\be
Z_{\rm SYM}^{\cN=(2,2)} = e^\Gamma.
\ee
To find the effective action we perform the standard
background-quantum splitting of the gauge superfield $V$
\cite{GGRS}
\be
e^{V} = e^{\Omega^\dag}e^{g\,v }e^\Omega\,,
\label{eV}
\ee
where $v$ is the so-called quantum gauge superfield while
$\Omega$ is a complex unconstrained prepotential which defines the
background gauge superfield $V_0$ as
\be
e^{V_0} = e^{\Omega^\dag}e^\Omega\,.
\ee
Upon this splitting the gauge symmetry is realized in two
different ways:
\bea
(i)&& e^\Omega \to e^{i\tau} e^\Omega\,,\qquad
 e^{g\,v} \to e^{i\tau} e^{g\, v} e^{-i\tau}\,,\label{tau}\\
(ii) && e^{\Omega} \to e^{i\lambda} e^\Omega
e^{-i\lambda}\,,\qquad
 e^{g\,v} \to e^{i\bar\lambda} e^{g\,v}e^{-i\lambda}\,.
\label{lambda}
\eea
Here $\tau$ and $\lambda$ are real and chiral
superfield parameters, respectively. The basic idea of the background field
method is to fix the gauge symmetry (\ref{lambda}) keeping the
invariance of the effective action under (\ref{tau}).

We will compute the one-loop effective action $\Gamma[V_0]$ for
the background gauge superfield $V_0$ taking values in the Cartan
subalgebra of the $u(N)$ gauge algebra
\be
V_0={\rm diag}(V_1,V_2,\ldots ,V_N)\,.
\label{Vdiag_}
\ee
Moreover, we assume that each of the superfields $V_I$ has
constant superfield strengths as in (\ref{G-const}). In
components, such a background is given by (\ref{comp-background})
for every $V_I$.

The one-loop effective action is defined by the action for
quadratic fluctuations around the chosen background. For arbitrary
background this action has a conventional form which is similar to
the $\cN=1$ $d=4$ \cite{GGRS} and $\cN=2$ $d=3$ \cite{my2} SYM
models
\be
S_2 = -\frac 12 \tr\int d^6 z \, E\, v(\nabla^\alpha \bar \nabla^2 \nabla_\alpha
 - 4i W^\alpha \nabla_\alpha) v\,.
\label{S2}
\ee
Here the superfield strength $W_\alpha$ and the gauge-covariant
derivatives $\nabla_\alpha$ and $\bar\nabla_\alpha$ are
constructed with the use of the background gauge superfield $V_0$
according to the rules (\ref{3.9}) and (\ref{GV}). Recall that the
background superfield $V_0$ corresponds to the constant scalar superfield
strengths $G$ and $H$ while the spinor superfield strengths
vanish, $W_\alpha=\bar W_\alpha=0$, see eq.\ (\ref{G-const}). For such a background the
action (\ref{S2}) simplifies to
\be
S_2 = -\frac12 \tr\int d^6z\, E \, v\nabla^\alpha\bar\nabla^2
\nabla_\alpha v\,.
\label{S2_}
\ee

The operator $\nabla^\alpha \bar\nabla^2 \nabla_\alpha$ in
(\ref{S2_}) is degenerate and requires gauge fixing. The gauge
symmetry under the $\lambda$-transformations (\ref{lambda}) is fixed
by imposing the standard conditions
\be
i\bar\nabla^2 v  = f\,,\qquad
i\nabla^2 v =\bar f\,,
\label{gauge-fixing}
\ee
where $f$ is a fixed covariantly chiral superfield, $\bar\nabla_\alpha
f=0$.

Following the standard procedure used for quantizing (superfield) gauge
theories \cite{GGRS}, one should introduce covariantly chiral ghost superfields $b$
and $c$, $\bar\nabla_\alpha b = \bar\nabla_\alpha c=0$.
The quadratic part of the ghost superfield action is
\be
S_{\rm FP} =\tr \int d^6z \, E \, (\bar b c- b\bar c)\,.
\label{FP}
\ee
Thus, the one-loop partition function gets the following
functional integral representation
\be
Z_{\rm SYM}^{\cN=(2,2)} =\int {\cal D}v {\cal D}b {\cal D}c {\cal
D}\varphi \,
\delta(f-i\bar\nabla^2 v) \delta(\bar f -i\nabla^2 v)
 e^{-S_2 - S_{\rm FP} }\,.
\label{Zfunc}
\ee
Upon averaging this expression with the weight
\be
1= \int {\cal D}f {\cal D}\varphi\, e^{\frac12 \tr \int d^6z\, E\,(\bar f f
+\bar \varphi \varphi)}\,,
\label{weight}
\ee
where $f$ and $\varphi$ are Grassmann-even and Grassmann-odd
chiral superfields, respectively, we end up with the gauge-fixing and
the Nielsen-Kallosh ghost superfield actions
\be
S_{\rm gf} = -\frac14 \tr\int d^6z \, E\, v\{ \nabla^2 ,
\bar\nabla^2 \} v\,,\qquad
S_\varphi = \frac12\tr \int d^6z \, E\, \bar\varphi \varphi\,.
\ee

The sum of the actions $S_{\rm gf}$ and $S_2$ can be recast as follows
\be
S_2 + S_{\rm gf} = -\tr\int d^6z \, E\, v\square_{\rm v} v\,,
\ee
where $\square_{\rm v}$ is the gauge-covariant Laplacian operator
acting in the space of general real superfields
\be
\square_{\rm v}=\frac1{4}\{\nabla^2,\bar\nabla^2 \}-\frac12
\nabla^\alpha \bar\nabla^2 \nabla_\alpha
+2i W^\alpha\nabla_\alpha\,.
\ee
Using the algebra of the covariant derivatives
(\ref{c-deriv}), for an arbitrary gauge superfield background
this operator can be written in the form
\bea
\square_{\rm v}&=&-\nabla^a\nabla_a
+(H+\frac1{r}M)^2
+(G-\frac i{2r}R)^2
+\frac1{2r}[\nabla^\alpha,\bar\nabla_\alpha]
\nn\\&&
+2i W^\alpha \nabla_\alpha
-2i \bar W^\alpha \bar\nabla_\alpha
-i(\nabla^\alpha W_\alpha)\,.
\label{cov-box}
\eea
In comparison with the three-dimensional case \cite{SS}, this
operator has additional term with the superfield strength $H$.

Recall that we consider the gauge superfield background
constrained by (\ref{G-const}). For such a background the form of
the operator (\ref{cov-box}) acting in the space of chargeless
scalar superfields simplifies to
\be
\square_{\rm v}=-\nabla^a\nabla_a
+H^2+G^2
+\frac1{2r}[\nabla^\alpha,\bar\nabla_\alpha]\,.
\ee

After averaging (\ref{Zfunc}) with the weight (\ref{weight}) the
integrals over all superfields become Gaussian and
we get the following form of the one-loop
partition function of the $\cN=(2,2)$ SYM model
\be
Z_{\rm SYM}^{\cN=(2,2)} ={\rm Det}^{-1/2}\square_{\rm v} \cdot
Z_\varphi\cdot Z_b\cdot Z_c\,,
\label{Znext}
\ee
where $Z_\varphi$, $Z_b$ and $Z_c$ are the one-loop partition
functions of the chiral ghost superfields.

Let us discuss the contribution to (\ref{Znext}) of the operator
$\square_{\rm v}$. In general, as a consequence of the gauge
invariance of the effective action, the trace of the logarithm of this
operator is given by a functional of superfield strengths $G$ and
$H$
\be
-\frac12 \Tr\ln \square_{\rm v} = \int d^6z \, E\, {\cal L}(G_I,
H_I)\,.
\label{expr}
\ee
As pointed out in (\ref{G-const}), we consider the constant superfield
strengths $G_I$ and $H_I$. Hence, the effective Lagrangian ${\cal L}(G_I,
H_I)$ is also a constant. Therefore the expression (\ref{expr}) is
proportional to the volume of the supercoset $\frac{SU(2|1)}{U(1)\times
U(1)}$ which vanishes according to (\ref{vanish-volume}). Thus,
the contribution from $\square_{\rm v}$ to (\ref{Znext}) is
trivial,
\be
{\rm Det}^{-1/2}\square_{\rm v} =1\,.
\label{Det1}
\ee
Note that this conclusion is completely analogous to the one for the
$\cN=2$ SYM model on $S^3$ \cite{SS}.

The equation (\ref{Det1}) shows that the partition function in the
$\cN=(2,2)$ SYM model receives contributions from the ghost
superfields only. These are Grassmann-odd chiral superfields in
the adjoint representation of the gauge group. It is important to
note that the R-charges of these superfields are
\be
q_{(b)}=q_{(c)}= 0\,,\qquad
q_{(\varphi)} =2\,.
\ee
Taking into account these values of the R-charges we apply the
formula (\ref{Zad}) to find the partition functions of the ghost
superfields
\bea
Z_\varphi^{-1} = Z_b = Z_c
=\prod_{I<J}^N\left(
\frac{(n_I-n_J)^2}{4}  + r^2 (\sigma_I - \sigma_J)^2
\right)\,.
\label{Zghosts}
\eea
Substituting these partition functions into (\ref{Znext}) and
taking into account (\ref{Det1}) we find
\be
Z_{\rm SYM}^{\cN=(2,2)}(\sigma_I,n_I)=\prod_{I<J}^N\left(
\frac{(n_I-n_J)^2}{4}  + r^2 (\sigma_I - \sigma_J)^2
\right)\,.
\label{parN22}
\ee
The one-loop partition function of the $\cN=(2,2)$ SYM model in this
form was obtained in \cite{Benini-Cremonesi,Gomes12} using
component field computations. Here we re-derived the same result
using the superfield method.

An interesting feature of the superfield approach for computing
the partition function in the $\cN=(2,2)$ SYM theory is that the
result (\ref{parN22}) appears solely due to the ghost superfields
(\ref{Zghosts}) while the gauge superfield itself does not
contribute (\ref{Det1}). At first sight this might seem strange
since in the ordinary component field computations
\cite{Benini-Cremonesi,Gomes12} there are non-trivial
contributions both from the ghosts and the fields from the
$\cN=(2,2)$ gauge supermultiplet. We stress that there is no
contradiction between the component field approach and the
superfield method since they give the same result. In fact, this
is not surprising because the details of computations depend
essentially on the gauge fixing condition. We use
the manifestly supersymmetric gauge fixing condition (\ref{gauge-fixing})
while the authors of \cite{Benini-Cremonesi,Gomes12} employed a
non-supersymmetric gauge.

\subsection{$\cN=(4,4)$ and $\cN=(8,8)$ SYM one-loop partition functions}

We will now compute the partition functions of the $\cN=(4,4)$ and $\cN=(8,8)$ SYM models on $S^2$ described by the actions
(\ref{SN=44}) and \eqref{N88}, respectively.

The action (\ref{SN=44}) depends on the parameter $q$ which is
associated with the R-charge of the chiral superfield $\Phi$ that
is part of the $\cN=(4,4)$ gauge supermultiplet. So, the partition
function of this model depends not only on the parameters of
the Coulomb branch, but also on $q$
\be
Z_{\rm SYM}^{\cN=(4,4)} =Z_{\rm SYM}^{\cN=(4,4)}(n_I,
\sigma_I;q)\,.
\ee
Here $n_I$ and $\sigma_I$ are the parameters which are related to the
vacuum values of the scalar fields of the vector multiplet $V$ as in
(\ref{comp-background}) and (\ref{G-const}). Note that in the
$\cN=(4,4)$ SYM model we can give vacuum values also to the
scalar fields in the chiral multiplet, $\phi_0= \langle \Phi
\rangle$, $\bar\phi_0=\langle \bar\Phi \rangle$. However, we
simplify the problem by considering vanishing values of these
scalars, $\phi_0 = \bar\phi_0 =0$, keeping in mind that the
dependence of the partition function on $\phi_0$ and
$\bar\phi_0$  can be easily restored by employing
the $SU(2)\sim SO(3)$ R-symmetry which rotates Re$\phi$, Im$\phi$ and $\sigma$.

In comparison with the $\cN=(2,2)$ SYM theory, the partition
function of the $\cN=(4,4)$ SYM receives also a contribution from
the chiral superfield $\Phi$,
\be
Z^{\cN=(4,4)}_{\rm SYM}={\rm Det}^{-1/2}(\square_{\rm v}
-\frac q{4r}[\nabla^\alpha,\bar\nabla_\alpha])\cdot Z_\varphi \cdot
Z_{b}\cdot Z_c\cdot Z_\Phi\,,
\label{Z-det-N4}
\ee
where $Z_\Phi$ is given in (\ref{Zad}). Note that in
(\ref{Z-det-N4}) the operator $\square_{\rm v}$ gets shifted by
the term $-\frac q{4r}[\nabla^\alpha,\bar\nabla_\alpha]$ which
originates from the second variational derivative of the CS-term
in (\ref{SN=44}). Applying
the same arguments as in (\ref{expr}) and (\ref{Det1}) to the
operator $\square_{\rm v}
-\frac q{4r}[\nabla^\alpha,\bar\nabla_\alpha]$ one can easily
argue that
\be
{\rm Det}^{-1/2}(\square_{\rm v}
-\frac q{4r}[\nabla^\alpha,\bar\nabla_\alpha]) =1\,.
\label{5.57}
\ee
Next, according to (\ref{Zghosts}), $ Z_\varphi$ and
$Z_{b}$ cancel against each other, and we end up with the following
expression for the $\cN=(4,4)$ SYM partition function
\be
Z^{\cN=(4,4)}_{\rm SYM}= Z_c\cdot Z_\Phi\,,
\label{Z-N4}
\ee
where the explicit expressions for $Z_c$ and $Z_\Phi$ are given in
(\ref{Zghosts}) and (\ref{Zad}), respectively.

In the end of section \ref{Sec411} we pointed out that the value
$q=0$ in the $\cN=(4,4)$ SYM model is singled out by the
requirement that the scalar fields should have non-negative
masses. For the vanishing R-charge the factors $Z_c$ and
$Z_\Phi$ in (\ref{Z-N4}) exactly cancel each other and the
partition function trivializes
\be
Z^{\cN=(4,4)}_{\rm SYM} |_{q=0} =1\,.
\ee
A similar trivialization of the one-loop partition function was also
observed in \cite{SS} for $\cN=4$ SYM on $S^3$.

The classical action of the $\cN=(8,8)$ SYM theory (\ref{N88})
contains the three chiral superfields $\Phi_i$ each of which has the fixed
R-charge $q=\frac 23$. Hence, the expression (\ref{Z-N4}) easily
generalizes to the case of $\cN=(8,8)$ SYM one-loop partition
function
\be
Z^{\cN=(8,8)}_{\rm SYM}= Z_c\cdot (Z_\Phi)^3|_{q=\frac 23}\,,
\label{Z-N8}
\ee
where the expression for $Z_c$ and $Z_\Phi$ are given in
(\ref{Zghosts}) and (\ref{Zad}), respectively.

\section{Localization}

\subsection{$\cN=(2,2)$ SYM partition function}
A representation for the partition function in a general
$\cN=(2,2)$ gauge theory which involves the gauge and chiral
multiplets was obtained in \cite{Benini-Cremonesi,Gomes12} using
the localization method for supersymmetric gauge theories. In this
section we discuss how the same representation can be
obtained using the superfield form of the $\cN=(2,2)$ SYM action.

Consider the $\cN=(2,2)$ SYM model (\ref{S-SYM1}) extended with
the FI-term (\ref{FI}),
\be
S=S_{\rm SYM}+S_{\rm FI}=-4\tr\int d^6z\, E\,\left(
\frac1{g^2}G^2-\xi V
\right)\,.
\label{SYM+FI}
\ee
In general \cite{Benini-Cremonesi,Gomes12}, one can also extend this action with the topological
term
\be
S_{\rm top}=-i\frac{\vartheta}{2\pi}\int \tr F\,,
\label{Stop}
\ee
where $F$ is a two-form field strength of the purely bosonic gauge field, $\tr F
=\tr dA$, such that  $\vartheta$  and the Fayet-Iliopoulos coupling constant $\xi$ form a
single complex coupling $\tau = \frac\vartheta{2\pi}+i\xi$.
However, we do not include this term in our consideration since
superspace formulation of the action (\ref{Stop}) is not known.

Before gauge fixing, the partition function in the model (\ref{SYM+FI}) is given by the
functional integral
\be
Z=\int {\cal D}V e^{-S_{\rm SYM}-S_{\rm FI}}\,.
\label{Z-SYM+FI}
\ee
In principle, $Z$ can depend on the both couplings $Z=Z(g^2,\xi)$.
However, standard localization arguments \cite{Pestun} can be used to demonstrate that
$Z$ is independent of $g^2$. Indeed, the $\cN=(2,2)$ SYM action is
known to be $Q$-exact with respect to a supersymmetry generator
$Q$ on $S^2$ \cite{Benini-Cremonesi,Gomes12}. Hence, one can
harmlessly deform the functional integral (\ref{Z-SYM+FI})
by introducing an arbitrary real parameter $t$,
\be
Z(t)=\int {\cal D}V e^{-tS_{\rm SYM}-S_{\rm FI}}\,,
\label{Z-tSYM+FI}
\ee
such that $Z$ is in fact independent of $t$, $\frac{d}{dt}Z(t)=0$,
and, hence, is independent of $g^2$ as well.
Owing to this property, we can compute the functional integral
(\ref{Z-tSYM+FI}) in the limit $t\to \infty$ where some
simplifications are expected. Indeed, at large $t$ the functional
integral localizes on the critical points $V_0$, i.e., on those field
configurations which are invariant under the supersymmetry and
for which the SYM action vanishes, $S_{\rm
SYM}[V_0]=0$. In superspace it is easy to find the
general solution of the latter equation
\be
S_{\rm SYM}=0\quad\Rightarrow\quad
W_\alpha=0\,,\quad G=G_0=const\,,\quad
H=H_0 = const\,.
\label{bg}
\ee
Indeed, the classical SYM action (\ref{S-SYM1}) is given by the
superfield Lagrangian proportional to $G^2$ or $H^2$ integrated
over the full superspace. However, according to
(\ref{vanish-volume}), such integrals vanish for constant
superfield strengths. Moreover, one can easily see that the
superfield background (\ref{bg}) is invariant under supersymmetry
variations on $S^2$ which have the general form
(\ref{general-transf}) since the superfields $G$ and $H$ are
neutral under the action of the generators $R$ and $M$. Therefore, in the superfield description,
the set of critical points is described by the constant scalar superfield
strengths.

For the gauge group $U(N)$, the constants $G_0$ and $H_0$ are
matrices in the Lie algebra $u(N)$. The standard arguments of
residual gauge invariance allow one to reduce the set of these
critical points to the Cartan subalgebra of the gauge algebra thus
leading to the appearance of the contribution of the Vandermonde determinant
into the path integral measure (see, e.g., \cite{BT} for a review). However, here we will achieve the
same result in a different way. We will show, a posteriori, that the correct
expression can be obtained by fixing the background gauge
superfield $V_0$ to belong to the Cartan subalgebra, by imposing this constraint on $V_0$ ``by hand''. In
this case the Vandermonde determinant contribution will appear
automatically as a part of the one-loop partition function of the
$\cN=(2,2)$ SYM theory. This procedure is, in fact, completely
analogous to the one given in \cite{SS} for the superfield gauge
theory on $S^3$, but we repeat its basic steps here for completeness.

Let us start by considering the gauge superfield background
(\ref{bg}) without additional restrictions. In the path integral
(\ref{Z-tSYM+FI}) we perform the background-quantum splitting
$V\to (V_0,\frac1{\sqrt t}v')$ similar to (\ref{eV}), but using the
parameter $t$ instead of the gauge coupling constant,
\be
e^V = e^{\Omega^\dag} e^{\frac1{\sqrt t}v'}e^\Omega\,,\qquad
e^{V_0}  = e^{\Omega^\dag}e^\Omega\,.
\label{Bg-q}
\ee
Upon this splitting we assume that the space of all the fields $\{V\}$ is a
direct sum of the spaces of the fields $\{ V_0\}$ and $\{v' \}$. Then,
the integration measure of the functional integral factorizes
\be
{\cal D}V = {\cal D}V_0 {\cal D}v'\,.
\label{DVDv}
\ee
This means that the modes which are taken into account by ${\cal
D}V_0$ should be absent in the measure ${\cal D}v'$. Recall that
the value of the gauge superfield $V_0$ is related to the constant
scalar gauge superfield strengths (\ref{bg}). Hence, in the
measure ${\cal D}v'$ the integration goes over such superfields
which have \emph{non-constant} superfield strengths. We denote the
space of these superfields by $\{v'\}$ to distinguish them from
the unconstrained superfields $\{ v \}$.

Following the same steps as in Section \ref{Sec5.2}, upon the
background-quantum splitting (\ref{Bg-q}) and fixing the gauge
freedom for the superfield $v'$ we get the following
representation of the path integral (\ref{Z-tSYM+FI})
\be
Z(t)=\int {\cal D}V_0 {\cal D}v' {\cal D}b {\cal D}c
\,\delta(f-i\bar\nabla^2 v') \delta(\bar f - i\nabla^2 v')\,
e^{-tS_{\rm SYM}[V_0,\frac1{\sqrt t}v']-S_{\rm FI}[V_0,\frac1{\sqrt t}v']
-S_{\rm FP}}\,.
\label{Z3}
\ee
The  part of the Faddeev-Popov action $S_{\rm FP}$ which is quadratic in superfields has the form
(\ref{FP}).

The basic idea of the localization method is to compute the
functional integral (\ref{Z3}) in the limit $t\to \infty$ in which
only quadratic fluctuations of the superfields around the
background $V_0$ survive,
\bea
-tS_{\rm SYM}[V_0,\frac1{\sqrt
t}v']&=&-S_2[V_0,v']+O(1/\sqrt t)\,,\nn\\
-S_{\rm FI}[V_0,\frac1{\sqrt t}v']&=&-S_{\rm FI}[V_0]+O(1/\sqrt
t)\,,
\eea
where the action $S_2[V_0,v']$ is given by (\ref{S2_}). Thus,
sending $t$ to infinity, we get the following representation for
the partition function (\ref{Z3})
\be
Z=\lim_{t\to\infty}Z(t)
=\int {\cal D}V_0\, e^{-S_{\rm FI}[V_0]}\cdot Z'_{\rm
SYM}\,,
\label{Z4}
\ee
where
\be
Z'_{\rm SYM}=\int {\cal D}v' {\cal D}b {\cal D}c {\cal
D}\varphi \,
\delta(f-i\bar\nabla^2 v') \delta(\bar f -i\nabla^2 v')
 e^{-S_2[V_0,v'] - S_{\rm FP} }
\label{Zfunc'}
\ee
is the one-loop $\cN=(2,2)$ SYM partition function which is very
similar to (\ref{Zfunc}), but with the restriction on the
superfields $v'$ such that they do not include the zero modes
corresponding to  the constant scalar superfield strengths. Recall that these
modes are taken into account by the measure ${\cal D}V_0$
according to (\ref{DVDv}).

With the use of the superfield methods it is difficult to compute the
functional integral (\ref{Zfunc'}) because of the
constraint on the integration values of the superfield $v'$. However, one can
rearrange the measure of the functional integral (\ref{Z4}) in such a way that
the integration over $v'$ becomes unconstrained. Recall that the
background superfield $V_0$ is the Lie-algebra-valued matrix
corresponding to the constant superfield strengths (\ref{bg}). This
matrix can be naturally decomposed as
\be
V_0 = V_0^{\mathfrak h}  + V_0^{\mathfrak r}\,,\qquad
V_0^{\mathfrak h} \in{\mathfrak h}\,,\quad
V_0^{\mathfrak r}\in {\mathfrak r}\,,
\ee
where the Lie algebra ${\mathfrak g}$ is given by the direct sum of the
Cartan subalgebra ${\mathfrak h}$ and the root space directions $\mathfrak
r$, ${\mathfrak g}={\mathfrak h}\oplus{\mathfrak r}$. Thus, the integration
measure ${\cal D}V_0$ decomposes as
\be
{\cal D}V_0  = {\cal D}V_0^{\mathfrak h}{\cal D}V_0^{\mathfrak r}\,.
\ee
Now, we combine the measures ${\cal D}V_0^{\mathfrak r}$ and ${\cal
D}v'$ together
\be
{\cal D}v = {\cal D}V_0^{\mathfrak r}{\cal D}v'
\ee
such that the new measure ${\cal D}v$ includes the missing zero modes of
fields $v'$ and the superfield $v$ becomes
unconstrained.\footnote{Note that without loss of generality the superfields
$v$ and $v'$ can be considered to belong to the space $\mathfrak r$ orthogonal to the Cartan subalgebra of
the gauge algebra since the corresponding Cartan components of these fields
do not interact with $V_0^{\mathfrak h}$.}
With this rearrangement of the integration measure in
(\ref{Zfunc'}) we end up with the following expression for the
partition function
\be
Z=\int {\cal D}V_0^{\mathfrak h} \, e^{-S_{\rm FI}[V_0^{\mathfrak h}]}\cdot
Z_{\rm SYM}[V_0^{\mathfrak h}]\,.
\label{Zfin}
\ee
In this expression the functional integration is performed over
the background superfield $V_0^{\mathfrak h}$ taking values in the
Cartan subalgebra of the gauge algebra and $Z_{\rm SYM}[V_0^{\mathfrak
h}]$ is precisely the $\cN=(2,2)$ SYM partition function (\ref{parN22}).

Note that we could arrive at the representation for the partition
function (\ref{Zfin}) by imposing the constraint on $V_0$ to
belong to the Cartan subalgebra from the very beginning. In this case we
do not need to care about the Vandermonde determinant contribution
to the functional integral because it is automatically taken into
account in $Z_{\rm SYM}[V_0^{\mathfrak h}]$.

For the gauge superfield background (\ref{Vdiag}), (\ref{G-const})
each of the superfields $V_I$ is given in components by
(\ref{comp-background}). Every $V_I$ in components
has just two degrees of freedom given by the real variable
$\sigma_I$ corresponding to the vacuum expectation value of the scalar $\sigma$ and
by  an integer $n_I$ which is related to the vacuum expectation value of another scalar $\eta$.
Thus, the integration measure of the functional integral
(\ref{Zfin}) can be rewritten as
\be
\int{\cal D}V_0^{\mathfrak h}\to   \int \prod_{I=1}^N d\sigma_I
\sum_{\mbox{\scriptsize all }n_I}
\label{reduct-measure}
\ee
In other words, one should integrate over all the continuous
parameters $\sigma_I$ and sum over all the integers $n_I$.

Using (\ref{FI-comp}), one cam bring the FI-term in the functional
integral (\ref{Zfin}) to the following form
\be
S_{\rm FI}[V_0]=\xi\int d^2x\,\sqrt h\, \tr D=
\frac{i\xi}{r}\sum_{I=1}^N \sigma_I {\rm Vol}(S^2)
=4\pi i r\xi\sum_{I=1}^N \sigma_I\,.
\label{6.17}
\ee
Finally, substituting (\ref{6.17}) and (\ref{parN22})
into (\ref{Zfin}) we end up with
\be
Z=\int \prod_{I=1}^N d\sigma_I
\sum_{\mbox{\scriptsize all }n_I} e^{-4\pi i r\xi \sum_{I=1}^N\sigma_I}
\prod_{I<J}^N\left(
\frac{(n_I-n_J)^2}{4}  + r^2 (\sigma_I - \sigma_J)^2
\right)\,.
\label{Zfin1}
\ee
In this form the partition function in the $\cN=(2,2)$
gauge theories was obtained in \cite{Benini-Cremonesi,Gomes12}
using component field methods for computing one-loop determinants.
Here we re-derived the same result starting with a superfield
formulation of this model.

Note also that, in general, the exponential of the topological
term (\ref{Stop}) can be inserted into the integral in
(\ref{Zfin1}), and also the contributions of chiral matter multiplets
can be taken into account. All these cases were studied in
\cite{Benini-Cremonesi,Gomes12}.

\subsection{Gaiotto-Witten and ABJ(M) models reduced to $S^2$}
The classical actions of Gaiotto-Witten (\ref{GW}) and ABJM models
(\ref{S-ABJM}) are very similar, so their partition functions can be constructed
using the same procedure which mimics the one for the corresponding three-dimensional
theories \cite{KWY,KWY1}. The essential difference among these
models is that the ABJM model has twice as many chiral
superfields that give extra contributions. Recall that we denote
the chiral superfield as $X_\pm$ while the gauge superfields are
$V$ and $\tilde V$. We consider the gauge group $U(M)\times U(N)$.

Before gauge fixing, the partition function in the Gaiotto-Witten
or ABJM models is represented by the functional integral
\be
Z=\int {\cal D}X_\pm {\cal D} V{\cal D}\tilde V
\, e^{-S[X,V,\tilde V]}\,,
\ee
where $S[X,V,\tilde V]$ is either $S_{\rm GW}$ or $S_{\rm ABJM}$. We deform this
partition function by inserting the $Q$-exact $\cN=(2,2)$ SYM action
(\ref{S-SYM1}) for the  both gauge superfields multiplied by a real parameter $t$
\be
Z(t) =\int {\cal D}X_\pm {\cal D} V{\cal D}\tilde V
\, e^{-S[X,V,\tilde V]-t S_{\rm SYM}[V]-tS_{\rm SYM}[\tilde V]}\,.
\label{Z-GW-ABJM}
\ee
For large $t$ the functional integral over the gauge
superfields localizes on the critical points $V_0$ and $\tilde V_0$ which are described by the
superfield equations (\ref{bg}) for each of the gauge superfields.
As has been explained in the previous subsection, we can further restrict
these superfields to belong to the Cartan subalgebra
\be
V_0={\rm diag}(V_1,V_2,\ldots,V_M)\,,\qquad
\tilde V_0={\rm diag}(\tilde V_1,\tilde V_2,\ldots,\tilde V_N)\,.
\ee
Each of $V_I$ and $\tilde V_J$ contains component fields
with values as in eq.\
(\ref{comp-background}), i.e., the background is described by the pairs
$(n_I,\sigma_I)$ and $(\tilde n_J,\tilde\sigma_J)$ corresponding
to vevs of the scalars in the vector multiplets.

Similar to (\ref{Bg-q}), we perform the background-quantum
splitting for $V$ and $\tilde V$ in (\ref{Z-GW-ABJM})
\be
V\to (V_0,\frac1{\sqrt t}v)\,,\qquad
\tilde V\to (\tilde V_0,\frac1{\sqrt t}\tilde v)\,.
\ee
For large $t$ it is sufficient to consider only quadratic
fluctuations in the SYM actions while the action $S[X,V,\tilde V]$
should be considered for purely background gauge superfields only,
\be
Z=\lim_{t\to \infty}Z(t) =\int {\cal D}X_\pm {\cal D} V_0 {\cal D}v{\cal D}\tilde
V_0 {\cal D}\tilde v
\, e^{-S[X,V_0,\tilde V_0]-S_2[V_0,v]-S_2[\tilde V_0,\tilde v]}\,,
\ee
where the action $S_2$ is given by (\ref{S2_}). Upon gauge fixing the transformations of superfields $v$
and $\tilde v$  in the standard way (\ref{gauge-fixing}), we get the following representation for the
partition function
\be
Z=\int {\cal D}V_0 {\cal D}\tilde V_0 \, e^{-S_{\rm CS}[V_0]+S_{\rm CS}[\tilde V_0]
-S_{\rm FI}[V_0 + \tilde V_0]}
\cdot Z_X\cdot Z_{\rm SYM}[V_0]\cdot Z_{\rm SYM}[\tilde V_0]\,,
\label{Z-GW-ABJM1}
\ee
where $Z_{\rm SYM}[V_0]$ and $Z_{\rm SYM}[\tilde V_0]$ have the form (\ref{parN22}) while $Z_X$ is the one-loop
partition function for the (anti)chiral superfields $X_\pm$.
The term $S_{\rm FI}[V_0 + \tilde V_0]$ is given by
(\ref{GW-FI}).

Recall that in the Gaiotto-Witten model we have one chiral superfield $X_+$ in the
bi-fundamental representation and another chiral scalar $X_-$ in the
anti-bi-fundamental representation while in the ABJM model the
number of these superfields is doubled. Hence, the one-loop
partition function $Z_X$ is a simple generalization of
(\ref{Zad}):
\be
Z_X
=\prod_{I,J}
\left[
\frac{\Gamma(\frac{q_+}2+\frac{|n_I-\tilde
n_J|}2+ir(\sigma_I-\tilde\sigma_J))
\Gamma(\frac{q_-}2+\frac{|n_I-\tilde n_J|}2-ir(\sigma_I-\tilde \sigma_J))}{
\Gamma(1-\frac{ q_+}2+\frac{|n_I-\tilde n_J|}2-ir(\sigma_I-\tilde\sigma_J))
\Gamma(1-\frac{ q_-}2+\frac{|n_I-\tilde n_J|}2+ir(\sigma_I-\tilde\sigma_J))}
\right]^p\,,
\label{Zx}
\ee
where $p=1$ for the Gaiotto-Witten model and $p=2$ for ABJM.

It is easy to find the values of the CS- and FI-terms in the
Gaiotto-Witten and ABJM actions in a form similar to eq.\ (\ref{6.17})
\bea
S_{\rm CS}[V_0]-S_{\rm CS}[\tilde V_0]&=&i\pi\kappa
\sum_{I=1}^M(\frac{n_I^2}{4}-\sigma_I^2 r^2)
-i\pi\kappa
\sum_{J=1}^N(\frac{\tilde n_J^2}{4}-\tilde\sigma_J^2 r^2)\,,\\
S_{\rm FI}[V_0+\tilde V_0]&=&-\frac{\pi}{4}\kappa r(q_+ - q_-)\left(\sum_{I=1}^M\sigma_I
+
\sum_{J=1}^N \tilde\sigma_J \right)\,.
\eea
We substitute these expressions into (\ref{Z-GW-ABJM1}) and take
into account that the functional measure reduces to conventional
integrations over all $\sigma_I$ and $\tilde\sigma_J$ and
sums over all $n_I$ and $\tilde n_J$ according to (\ref{reduct-measure})
\bea
Z&=&  \int \prod_{I=1}^M  d\sigma_I  \sum_{\mbox{\scriptsize all }n_I}
 \int \prod_{J=1}^N d\tilde \sigma_J \sum_{\mbox{\scriptsize all }\tilde n_J}
Z_X\cdot Z_{\rm SYM}^{\cN=(2,2)}(\sigma_I,n_I)\cdot
Z_{\rm SYM}^{\cN=(2,2)}(\tilde\sigma_J,\tilde n_J)
 \nn\\&&
\times
\exp\left\{
-i\pi \kappa\sum_{I=1}^M(\frac{n_I^2}{4}-\sigma_I^2 r^2+\frac i4(q_+-q_-)r\sigma_I)
\right\}
\nn\\&&
\times\exp\left\{
i\pi\kappa\sum_{J=1}^N(\frac{\tilde n_J^2}{4}-\tilde\sigma_J^2 r^2-\frac i4(q_+-q_-)r\tilde\sigma_J)
\right\}\,.
\label{5.89}
\eea
Here $Z_X$ and $Z_{\rm SYM}^{\cN=(2,2)}(\sigma_I,n_I)$
are given by (\ref{Zx}) and (\ref{parN22}), respectively.

By shifting the integration variables
\be
\sigma_I \to \sigma_I-\frac i{8r}(q_+-q_-)\,,\qquad
\tilde\sigma_J \to \tilde\sigma_J+\frac i{8r}(q_+-q_-)
\ee
the expression (\ref{5.89}) can be slightly simplified
\bea
Z&=&  \int \prod_{I=1}^M  d\sigma_I  \sum_{\mbox{\scriptsize all }n_I}
 \int \prod_{J=1}^N d\tilde \sigma_J \sum_{\mbox{\scriptsize all }\tilde n_J}
 Z_{\rm SYM}^{\cN=(2,2)}(\sigma_I,n_I)\cdot
Z_{\rm SYM}^{\cN=(2,2)}(\tilde\sigma_J,\tilde n_J)
\label{5.91}\\&&
\times
\exp\{
-i\pi \kappa\sum_{I=1}^M(\frac{n_I^2}{4}-\sigma_I^2 r^2)
+i\pi\kappa \sum_{J=1}^N(\frac{\tilde n_J^2}{4}-\tilde\sigma_J^2 r^2)
+\frac{i\pi\kappa}{16}(N-M)(q_+ - q_-)^2\}
\nn\\&&\times
\prod_{I,J}
\left[
\frac{\Gamma(\frac{q_++q_-}4+\frac{|n_I-\tilde
n_J|}2+ir(\sigma_I-\tilde\sigma_J))
\Gamma(\frac{q_++q_-}4+\frac{|n_I-\tilde n_J|}2-ir(\sigma_I-\tilde \sigma_J))}{
\Gamma(1-\frac{ q_++q_-}4+\frac{|n_I-\tilde n_J|}2-ir(\sigma_I-\tilde\sigma_J))
\Gamma(1-\frac{q_++ q_-}4+\frac{|n_I-\tilde n_J|}2+ir(\sigma_I-\tilde\sigma_J))}
\right]^p\,.\nn
\eea
One can see that when the ranks of the gauge groups are equal, $M=N$,
the charges of the chiral superfields $q_+$ and $q_-$ enter the
partition function in the combination $q_++q_-$. Therefore, for $M=N$
without loss of generality we can assume that
\be
q:= q_+ = q_-\,.
\ee
Another important observation is that upon re-scaling the integration
variables $\sigma_I\to \frac1r\sigma_I$ and $\tilde\sigma_J\to
\frac1r\tilde\sigma_J$ the partition function becomes independent
of the radius of the sphere. So, putting for simplicity $r=1$ and
$M=N$, the partition functions of the Gaiotto-Witten ($p=1$) and ABJM
($p=2$) models which follow from eq.\
(\ref{5.91}) get the following explicit form
\bea
Z_{\rm GW}&=&\int (\prod   d\sigma_I)  \sum_{\mbox{\scriptsize all }n_I}
 \int (\prod d\tilde \sigma_J) \sum_{\mbox{\scriptsize all }\tilde n_J}
\nn\\&&
\times
\exp\left\{
-i\pi \kappa\sum (\frac{n_I^2}{4}-\frac{\tilde n_I^2}{4}-\sigma_I^2 +\tilde\sigma_I^2)\right\}
\nn\\&&\times
\prod_{I<J}\left(
\frac{(n_I-n_J)^2}{4}  +  (\sigma_I - \sigma_J)^2
\right)
\prod_{I<J}\left(
\frac{(\tilde n_I-\tilde n_J)^2}{4}  + (\tilde\sigma_I - \tilde\sigma_J)^2
\right)\\&&
\times\prod_{I,J}
\frac{\Gamma(\frac{q}2+\frac{|n_I-\tilde
n_J|}2+i(\sigma_I-\tilde\sigma_J))
\Gamma(\frac{q}2+\frac{|n_I-\tilde n_J|}2-i(\sigma_I-\tilde \sigma_J))}{
\Gamma(1-\frac{q}2+\frac{|n_I-\tilde n_J|}2-i(\sigma_I-\tilde\sigma_J))
\Gamma(1-\frac{q}2+\frac{|n_I-\tilde
n_J|}2+i(\sigma_I-\tilde\sigma_J))}\,,\nn\\
Z_{\rm ABJM}&=&\int (\prod  d\sigma_I)  \sum_{\mbox{\scriptsize all }n_I}
 \int (\prod d\tilde \sigma_J) \sum_{\mbox{\scriptsize all }\tilde n_J}
\nn\\&&
\times
\exp\left\{
-i\pi \kappa\sum (\frac{n_I^2}{4}-\frac{\tilde n_I^2}{4}-\sigma_I^2 +\tilde\sigma_I^2)\right\}
\nn\\&&\times
\prod_{I<J}\left(
\frac{(n_I-n_J)^2}{4}  +  (\sigma_I - \sigma_J)^2
\right)
\prod_{I<J}\left(
\frac{(\tilde n_I-\tilde n_J)^2}{4}  + (\tilde\sigma_I - \tilde\sigma_J)^2
\right)\nn\\&&
\times\prod_{I,J}
\left[
\frac{\Gamma(\frac{1}4+\frac{|n_I-\tilde
n_J|}2+i(\sigma_I-\tilde\sigma_J))
\Gamma(\frac{1}4+\frac{|n_I-\tilde n_J|}2-i(\sigma_I-\tilde \sigma_J))}{
\Gamma(\frac34+\frac{|n_I-\tilde n_J|}2-i(\sigma_I-\tilde\sigma_J))
\Gamma(\frac34+\frac{|n_I-\tilde n_J|}2+i(\sigma_I-\tilde\sigma_J))}
\right]^2\,. \label{ABJMZ}
\eea
In \eqref{ABJMZ}  we have also taken into account that the R-charges of chiral
superfields in the ABJM model are fixed to be $q_+=q_- = \frac12$.

\section{Discussion}

To summarize, in this paper we have elaborated on a superfield
approach based on the supercoset $\frac{SU(2|1)}{U(1)\times U(1)}$
for studying classical and quantum aspects of supersymmetric field
theories on $S^2$. We constructed the supersymmetric Cartan forms,
supercurvature, supertorsion and supercovariant derivatives on
this coset and applied them for constructing classical
actions for gauge and chiral superfields.

We have also given classical actions for various models with extended
supersymmetry on $S^2$ in terms of the $\cN=(2,2)$ superfields.
Among them, there are the actions for the $\cN=(4,4)$ hypermultiplet, $\cN=(4,4)$ and $\cN=(8,8)$ SYM models as well as
the actions for the  Gaiotto-Witten and ABJM theories reduced to $S^2$. For all these models
we have derived the transformations
of hidden supersymmetries realized on the $\cN=(2,2)$ superfields.
To the best of our knowledge, the classical superfield actions for the models
with extended supersymmetry on $S^2$ have not been considered before.

We have demonstrated that the superfield method facilitates the computation of the partition functions of
supersymmetric gauge and matter theories on $S^2$ and helps finding critical points in the space of fields for the localization technique. In particular, we have re-derived
the known expressions for the one-loop partition functions found
originally in \cite{Benini-Cremonesi,Gomes12} for the $\cN=(2,2)$
SYM and the chiral superfield models. An advantage of the superfield
method is that the cancellations among bosonic and
fermionic contributions to the one-loop determinants of the quadratic
fluctuations occur automatically. We have also demonstrated how the localization technique
applies to the  superfield description of the $\cN=(2,2) $ SYM model which was
originally considered in \cite{Benini-Cremonesi,Gomes12}. A new
result of this paper is the expression for the partition functions of the
Gaiotto-Witten and ABJM models reduced to $S^2$. For these
models the localization formula is very similar to the one for the
corresponding models on $S^3$ \cite{KWY,KWY1} and differs from it
mainly by the form of the one-loop determinants for the chiral and gauge multiplets. It
would be instructive to study the large $N$ behavior of
the partition function in the ABJM model reduced to $S^2$ and compare
it with the corresponding $S^3$ partition function \cite{Drukker}. It would also be of interest to elaborate on peculiarities of the superconformal structure of the $S^2$--counterparts of the Gaiotto--Witten and ABJM models in comparison with the $S^3$ case
\cite{SS,Kuzenko:2014yia}.

It would be very natural to extend
our results to the superfield models in higher--dimensional ($d\geq
4$) curved backgrounds
\footnote{Quantum mechanical (i.e. $d=1$) models on
different cosets of $SU(2|1)$ were considered in
\cite{Ivanov:2013ova,Ivanov:2013cea}.}. However, already in $d=4$ the minimal supersymmetry on the four-sphere is $\cN=2$, and it is well known that the quantization of
$\cN=2$ SYM and hypermultiplets keeping supersymmetry off-shell requires special methods
such as the use of harmonic superspace \cite{GIKOS,GIOS1,GIOS2,HSS}. It is very tempting to
extend harmonic superspace techniques to the case of superfield models on the sphere
or in the $AdS$ space.

Another possible extension of the results of this paper
could be the consideration of  twisted chiral and vector $\cN=(2,2)$ supermultiplets.
As was demonstrated in \cite{Kaehler0,Kaehler1},
quantum partition functions of such models compute the exact
K\"ahler potential for Calabi-Yau target space of $\cN=(2,2)$ non-linear
sigma-models. In
superspace, classical actions for these models were
systematically studied in \cite{CC}. It would be of interest to
develop a superfield approach for computing partition functions of
these models. This
issue becomes even more intriguing for the two-dimensional models
with extended supersymmetry on $S^2$. Indeed, as was pointed out
in earlier papers \cite{GI,Gates1995,I}, there are many inequivalent versions of twisted
multiplets with $(4,4)$ supersymmetry in flat superspace. Assuming
that these models allow for a superfield description in the curved
superspace based on the supercoset $\frac{SU(2|1)}{U(1)\times U(1)}$, it
is tempting to understand the difference among these models on the
quantum level by comparing their partition functions. These
problems require a separate systematic study.

The papers \cite{Benini-Cremonesi,Gomes12} showed that the
two-dimensional supersymmetric theories exhibit rich quantum
dynamics with many non-trivial dualities. This motivates further
study of low-energy dynamics of these models and, in particular,
their low-energy effective actions. Note that the low-energy
effective actions in three-dimensional gauge and matter theories in the
flat $\cN=2$ superspace were derived in \cite{my2,my1,my3,my4,my5,my6}.

\vspace{5mm} {\bf Acknowledgements.} The authors wish to thank
Jaume Gomis for the suggestion to look at the superfield
description of field theories on curved supermanifolds and their
localization. This work was
partially supported by the Padova University Project CPDA119349
and by the INFN Special Initiative ST\&FI. Work of I.B.S. was also
supported by the RFBR grants
Nr.\ 12-02-00121, 13-02-90430 and 13-02-91330 and by the LRSS grant Nr.\
88.2014.2.


\end{document}